\documentclass[11pt,reqno]{amsart}

\usepackage[hmargin=0.8in,twosideshift=0in,height=8.6in]{geometry}
\usepackage{amssymb,amsthm}
\usepackage{delarray,verbatim}
\usepackage{natbib}
\renewcommand{\cite}{\citet}

\usepackage{ifpdf}
\ifpdf
\usepackage[pdftex]{graphicx}
\else
\usepackage[dvips]{graphicx}
\fi

\usepackage{ifpdf}
\usepackage{color}
\definecolor{webgreen}{rgb}{0,.5,0}
\definecolor{webbrown}{rgb}{.8,0,0}
\definecolor{emphcolor}{rgb}{0.95,0.95,0.95}

\usepackage{hyperref}
\hypersetup{%
          colorlinks=true,
          linkcolor=webbrown,
          filecolor=webbrown,
          citecolor=webgreen,
          breaklinks=true}
\ifpdf
\hypersetup{pdftex,
            pdfstartview=FitH, 
            bookmarksopen=true,
            bookmarksnumbered=true
}
\else
\hypersetup{dvips}
\fi

\numberwithin{equation}{section} 

\newtheorem {thm}{Theorem}[section]
\newtheorem {prop}[thm]{Proposition}
\newtheorem {lem}[thm]{Lemma}

\newtheorem {cor}[thm]{Corollary}

\newtheorem{defi}{Definition}[section]

\theoremstyle{remark}

\newcommand{\noi}{\noindent}
\newcommand{\la}{\lambda}
\renewcommand{\a}{\alpha}
\newcommand{\E}{\mathbb{E}}
\renewcommand{\P}{\mathbb{P}}
\newcommand{\R}{{\bf R}}
\newcommand{\rp}{{\bf R}_+}
\newcommand{\zp}{{\bf Z}_+}

\newcommand{\eps}{\varepsilon}
\newcommand{\pf}{\begin{proof}}

\def\ba{\begin{array}}
\def\ea{\end{array}}
\def\beq{\begin{equation}}
\def\endeq{\end{equation}}
\def\bes{\begin{equation*}}
\def\ees{\end{equation*}}
\def\bea{\begin{eqnarray}}
\def\eea{\end{eqnarray}}
\def\beas{\begin{eqnarray*}}
\def\eeas{\end{eqnarray*}}
\def\bi{\begin{itemize}}
\def\ei{\end{itemize}}

\title[Regularity of the Minimal Probability of Ruin]{Proving Regularity of the Minimal Probability of Ruin \\ via a Game of Stopping and Control } \thanks{This version: August 27, 2010}
\thanks{We are very grateful to the anonymous Associate Editor and the referees for their incisive comments. Their comments helped us improve our paper in significant ways. We also would like to thank Huy\^{e}n Pham for his feedback.}

\author{Erhan Bayraktar }
\address[E. Bayraktar]{Department of
  Mathematics, University of Michigan, Ann Arbor, MI 48109}
\email{erhan@umich.edu}
\thanks{E. Bayraktar is supported in part by the National Science Foundation by an applied mathematics research grant, DMS-0906257, and a CAREER grant, DMS-0955463, and in part by the Susan M. Smith Professorship.}

\author{Virginia R. Young}
\address[V. R. Young]{Department of
  Mathematics, University of Michigan, Ann Arbor, MI 48109}
\email{vryoung@umich.edu}
\thanks{V. R. Young is supported in part by the Cecil J. and Ethel M. Nesbitt Professorship.}

\date{}

\begin{document}
\begin{abstract}
We reveal  an interesting convex duality relationship between two problems: (a) minimizing the probability of lifetime ruin when the rate of consumption is stochastic and when the individual can invest in a Black-Scholes financial market;  (b)  a controller-and-stopper problem, in which the controller controls the drift and volatility of a process in order to maximize a running reward based on that process,  and the stopper chooses the time to stop the running reward and rewards the controller a final amount at that time.  Our primary goal is to show that the minimal probability of ruin, whose stochastic representation does not have a classical form as does the utility maximization problem (i.e., the objective's dependence on the initial values of the state variables is implicit), is the unique classical solution of its Hamilton-Jacobi-Bellman  (HJB) equation, which is a non-linear boundary-value problem.  We establish our goal by exploiting the convex duality relationship between (a) and (b).

\noi {\bf MSC 2000 Classification:} Primary 93E20, 91B28; Secondary 60G40.  

\medskip

\noi {\bf JEL Classification:} Primary G11; Secondary C61.

\medskip


\end{abstract}

\maketitle

\section{Introduction and Motivation}\label{sec:1}

The main goal of this paper is to prove regularity of the minimum probability of lifetime ruin when the rate of consumption is stochastic and the individual invests in a Black-Scholes market to cover her consumption needs. We will refer to this optimization problem as Problem 1.  The Hamilton-Jacobi-Bellman (HJB) equation corresponding to this problem is a boundary-value problem. {\it A priori}  regularity of this function is not clear, unlike the value functions corresponding to utility maximization problems, since the dependence of the objective function on the initial values of the state variables is implicit.  (In this paper, we call a function {\it regular} when it is convex/concave and is the classical solution of the corresponding HJB equation.)  As a first step, we reduce the dimension of this problem. The resulting problem, which we will refer to as Problem 2, surprisingly has also an economic meaning: It is the problem of minimizing the probability of lifetime ruin for which the individual has constant consumption, with the opportunity set to cover her consumption consisting of two risky assets. 

Next, we consider a controller-and-stopper game, which we will refer to as Problem 3. The analysis of Problem 3 is crucial in the proof of the regularity of the minimum probability of ruin. In this game, first, the controller controls the drift and volatility of a process in order to maximize a running reward based on that process; then, the stopper chooses the time to stop the running reward and rewards the controller a final amount at that time.  We extensively analyze this problem, considering this problem not only as an intermediary tool but as an interesting problem in its own right.  In particular, we show that the value function is concave and is a classical solution of the corresponding HJB equation.  In fact, we consider a sequence of controller-and-stopper problems parameterized by the pay-off  function.  By taking the convex dual of each element, we obtain a sequence of convex, regular functions that uniformly converges to the minimum probability of ruin.  This fact leads to the proof of regularity the minimum probability of ruin in Problem 2, which in turn leads to regularity of Problem 1.  Please see Section~\ref{sec:3.1} for a detailed outline of our proof of regularity of the value functions of Problems 1, 2, and 3, and in particular, for the ordering of their regularity proofs.

When an individual determines an optimal investment policy, the resulting optimal policy depends on the criterion used.  Young (2004) proposes minimizing the probability of ruin as an alternative criterion to maximizing one's expected discounted utility of consumption and bequest.  Minimizing the probability that one outlives one's wealth is arguably an ``objective'' goal as compared with the goal of maximizing utility, in which one has to specify a ``subjective'' utility function.  For further motivation of this problem, see Bayraktar and Young (2007a,b), Browne (1995), Milevsky and Robinson (2000), Milevsky, Ho, and Robinson (1997), and Milevsky, Moore, and Young (2006).

In the first of the two ruin minimization problems, we assume that the individual can invest in a financial market with one risky and one riskless asset.  Young (2004) considers this problem in the case for which consumption is either constant or a constant proportion of wealth.  In this paper, we assume that the individual consumes at a rate that follows a diffusion that is correlated with the risky asset's price process.  It is important to consider random consumption because even though consumption is to some extent under the control of the individual, pressure from inflation or unexpected events can cause even the most frugal of individuals to experience random required consumption. Note that the solutions of the first two problems are useful in deriving mutual fund theorems for which the optimization criterion is the probability of ruin. See the note by Bayraktar and Young (2008).

Games of stopping and control were recently studied by Karatzas and Sudderth (2001).  They study a zero-sum game for which the controller selects the coefficients of a linear diffusion  on a given interval, while the stopper can halt the process at any time.  The arguments in this paper work when there is no running reward.  Karatzas and Zamfirescu (2006, 2008), on the other hand, develop a martingale approach for studying controller-and-stopper games when only the drift can be controlled, and they find conditions under which the game has a value. More recently, Buckdahn and Li (2009) considered controller and stopper games in a very general framework and analyzed the viscosity solution property of the value functions.

The remainder of the paper is organized as follows:  In Section~\ref{sec:2}, we present the three control problems along with our major theorems.  Section~\ref{sec:3} is home to our proofs.  Here, we see how  regularity of the controller-and-stopper problem leads to regularity of the minimum lifetime ruin probability.  Please see Section~\ref{sec:3.1} for the outline of the proofs given in this section.  Section~\ref{sec:4} concludes the paper.

\section{Three Related Optimal Control Problems}\label{sec:2}

In this section, we describe three related optimal control problems.  In Section~\ref{sec:2.1}, we consider our main  problem, the problem of minimizing the  probability of lifetime ruin when the rate of consumption is stochastic and when the individual can invest in a Black-Scholes financial market.  In Section~\ref{sec:2.2}, we consider the problem of minimizing the probability of lifetime ruin when the rate of consumption is constant but the individual can invest in two risky correlated assets.  In Section~\ref{sec:2.3}, we consider a controller-and-stopper problem.  First, the controller controls the drift and volatility of a process  in order to maximize a running reward based on that process.  Then, the stopper chooses the time to stop the running reward and pays the controller a final amount at that time.  This final amount is a function of the value of the process at the time of stopping.  In Section~\ref{sec:2.4}, we show how the three control problems described in Sections \ref{sec:2.1} through \ref{sec:2.3} are related: The second problem is obtained from the first problem after reducing the dimension, and the third problem is the concave dual of the second.  This last relationship is crucial since it is not clear {\it a priori} that the value functions of the first two problems are convex or smooth, and in Section~\ref{sec:3} we heavily rely on this relationship in our proof to show the regularity of these value functions. 

One should note that  in this section we merely state the problems, and their regularity results, and we summarize the proofs and give the relationship among the three problems. It will be clear in Section~\ref{sec:3} (please see the outline in Section~\ref{sec:3.1}), how the regularity of the third problem leads to the regularity of the second, which in turn leads to the regularity of the first.

\subsection{Probability of Lifetime Ruin with Stochastic
Consumption}\label{sec:2.1}

In this section, we present the financial ingredients that affect the individual's wealth, namely, random consumption, a riskless asset, and a risky asset.  We assume that the individual invests in order to minimize the probability that her wealth reaches zero before she dies.

Let $(\Omega, \mathcal{F}, \{\mathcal{F}_t\}_{t \geq 0}, \mathbb{P})$ be a filtered probability space that supports two standard Brownian motions $B^c$ and $B^S$, whose correlation coefficient is given by $\rho \in (-1, 1)$. 
The individual consumes at a random continuous rate $c_t$ at time $t$.  One can interpret this consumption rate as the {\it net} consumption rate offset by (possibly random) income. We assume that $c_t$ follows geometric Brownian motion given
by
\[dc_t = c_t(a \, dt + b \, dB^c_t), \quad c_0 = c > 0,\]
\noi in which $b > 0$. The individual invests in a riskless asset whose price at time $t$, $X_t$, follows the deterministic process $dX_t = rX_t dt, X_0 = x > 0$, for some fixed rate of interest $r > 0$.  Also, the individual invests in a risky asset whose price at time $t$,
$S_t$, follows geometric Brownian motion given by
\[
dS_t = S_t(\mu \, dt + \sigma \, dB^S_t), \quad S_0 = S > 0,
\]
\noi in which $\sigma > 0$. Note that we preclude $|\rho| = 1$ because one can explicitly compute the value function in that case, as in Young (2004). 

Let $W_t$ be the wealth at time $t$ of the individual, and let $\pi_t$ be the amount that the decision maker invests in the risky asset at that time.  It follows that the amount invested in the riskless asset is $W_t - \pi_t$, and wealth follows the process
\[
dW_t = \left( r \, W_t + (\mu - r) \, \pi_t  - c_t \right) dt + \sigma \, \pi_t \, dB^S_t, \quad W_0 = w > 0.\]

Define a hitting time $\tau_0$ associated with the wealth process by $\tau^{w,c}_0 = \inf \{  t\ge 0: W_t \le 0 \}$.  This hitting time is the time of ruin.  Also, define the random time of death of the individual by $\tau_d$.  We assume that $\tau_d$ is exponentially distributed with parameter $\la$ (that is, with expected time until death equal to $1/\la$); this parameter is also known as the {\it hazard rate} of the individual.  Even though we assume that $\tau_d$ is exponentially distributed (which is equivalent to a constant hazard rate), all our results extend to the case for which the hazard rate is time dependent.  However, we only consider a constant hazard rate in this paper to simplify the presentation.  We assume that $\tau_d$ is independent of the $\sigma$-algebra generated by the Brownian motions $B^c$ and $B^S$.

By {\it probability of lifetime ruin}, we mean the probability that wealth reaches 0 before the individual dies, that is, $\tau^{w,c}_0 < \tau_d$.   We minimize with respect to the set of admissible investment strategies $\mathcal{A}$, which is a collection of  ${\{ \mathcal{F}_t \}}$-progressively measurable strategies $\pi$
 (in which $\mathcal{F}_t$ is the augmentation of $\sigma(B^c_s, B^S_s: 0 \le s \le t)$)  that satisfy the integrability condition $\int_0^t \pi_s^2 \, ds < \infty$, almost surely, for all $t \ge 0$.  
The minimum probability of lifetime ruin $\psi$ is given by
\beq \label{eq:2-1}
\psi(w, c) = \inf_{\pi \in \mathcal{A}} \mathbb{P} \left( \tau_0^{w,c} < \tau_d \right).
\endeq
 We have the following theorem for the minimum probability of lifetime ruin $\psi$,  which is the main result of the paper.

\begin{thm}\label{thm:2.1}
The minimum probability of lifetime ruin $\psi$ given in $\eqref{eq:2-1}$ is strictly decreasing and strictly convex with respect to $w$, strictly increasing with respect to $c$, and lies in $\mathcal{C}^2(\rp^2)$.  Additionally, $\psi$ is the unique solution of the following
Hamilton-Jacobi-Bellman $($HJB$\, )$ equation on $\rp^2:$
\beq \label{eq:2.2}
\begin{split}
  &\la \, v = (rw - c) \, v_w + a \, c \, v_c + {1 \over 2} \, b^2 \, c^2 \, v_{cc} + \min_\pi \left[ (\mu - r) \, \pi \, v_w + {1 \over 2} \, \sigma^2 \, \pi^2 \, v_{ww} + \sigma \, \pi \, b \, c \, \rho \, v_{wc} \right], \\
&  v(0, c) = 1 \hbox{ and } v(w, 0) = 0.
\end{split}
\endeq
The optimal investment strategy $\pi^*$ is given in feedback form by
\beq \label{eq:moptinvst}
\pi^*_t = -{(\mu - r) \, \psi_w \, (W^*_t, c_t) + \sigma \, b \, \rho \, c_t \, \psi_{wc}(W^*_t, c_t) \over \sigma^2 \, \psi_{ww}(W^*_t, c_t)},
\endeq
in which $W^*$ is the optimally controlled wealth process.
\end{thm}
\begin{proof}
 See Section~\ref{sec:3.1}, item 11, for an outline of the proof of this theorem, and see Section~\ref{sec:3.4} for the proof itself.  
\end{proof}
Let us comment on the proof of this theorem: It is far from clear that $\psi$ is convex or smooth.  As a first step, we reduce the dimension of the problem from two variables to one (and obtain the problem given in the next section).  We, then, construct a regular sequence of convex functions that converges uniformly to the value function that we obtain after the dimension reduction.  We construct this sequence by taking the Legendre transform of the controller-and-stopper problem we introduce in Section~\ref{sec:2.3}.  The regularity analysis of the stopper-controller problem in Section~\ref{sec:2.3} turns out to be simpler, which, in turn, provides us a useful means for proving regularity of $\psi$.

\subsection{Probability of Lifetime Ruin with Two Risky Assets}\label{sec:2.2}

Consider two (risky) assets with prices $\tilde S^{(1)}$ and $\tilde S^{(2)}$ following the diffusions

$$d \tilde S^{(1)}_t = \tilde S^{(1)}_t \left( \tilde r \, dt + b \sqrt{1 - \rho^2} \, d \tilde B^{(1)}_t \right),
\quad d \tilde S^{(2)}_t = \tilde S^{(2)}_t \left( \tilde \mu \, dt + \sqrt{b^2(1 - \rho^2) + \sigma^2} \, d \tilde B^{(2)}_t \right),$$

\noi in which $\tilde r = r - a + b^2 + (\mu - r - \sigma b \rho) \rho b/\sigma$ and $\tilde \mu = \mu - r -\sigma b \rho + \tilde
r$.  Also, $\tilde B^{(1)}$ and $\tilde B^{(2)}$ are correlated standard Brownian motions on a probability space $(\widetilde{\Omega},\widetilde{\mathcal{F}},\widetilde{\P})$ with correlation coefficient

$$\tilde \rho = {b \sqrt{1 - \rho^2} \over \sqrt{b^2(1 - \rho^2) + \sigma^2}}.$$

Suppose an individual has wealth $Z_t$ at time $t$, consumes at the constant rate of 1, and wishes to invest in these two assets in order to minimize her probability of lifetime ruin.  Let $\tilde \pi_t$ be the dollar amount that the individual invests in the second asset at time $t$; then, $Z_t - \tilde \pi_t$ is the amount invested in the first asset at time $t$.

It follows that the wealth process $Z$ follows the dynamics
\beq \label{eq:Z-dyn}
\begin{split}
&d Z_t = - dt + \left( Z_t - \tilde \pi_t \right) \left( \tilde r \, dt + b \sqrt{1 - \rho^2} \, d \tilde B^{(1)}_t \right) + \tilde \pi_t \left( \tilde \mu \, dt + \sqrt{b^2(1 - \rho^2) + \sigma^2} \, d \tilde B^{(2)}_t \right) \\
&= \left( (\tilde r Z_t - 1) + (\mu - r - \sigma b \rho) \tilde \pi_t \right) dt + Z_t \, b \, \sqrt{1 - \rho^2} \, d \tilde B^{(1)}_t  + \tilde \pi_t \left( \sqrt{b^2(1 - \rho^2) + \sigma^2} \, d \tilde B^{(2)}_t - b \sqrt{1 - \rho^2} \, d \tilde B^{(1)}_t \right)
\end{split}
\endeq
\noi with $Z_0 = z$. Now, denote minimum probability of lifetime ruin for this individual by $\phi$.  Specifically, define $\phi$ by
\beq \label{eq:2.4}
\phi(z) = \inf_{\tilde \pi \in \widetilde{\mathcal{A}}}\widetilde{\P} \left( \tilde \tau_0^{z} < \tau_d \right), 
\endeq
in which $\tilde \tau_0^z = \inf \{ t \ge 0: Z_t \le 0  \}$ is the time of ruin.  Also, $\widetilde{\mathcal{A}}$ is the set of admissible strategies for this problem, defined similarly as we defined $\mathcal{A}$.

 Although it is not obvious, it turns out that $\psi(w, c) = \phi(w/c)$, as we will show later in Section~\ref{sec:2.4}; therefore, $\phi$ arises by reducing the dimension of $\psi$.   It is remarkable that $\phi$ {\it itself} is the minimum probability of ruin for a problem that has economic meaning. We have the following theorem for the minimum probability of lifetime ruin $\phi$.

\begin{thm}\label{thm:2.2}
The minimum probability of lifetime ruin $\phi$ given in $\eqref{eq:2.4}$ is strictly decreasing, strictly convex, and $\mathcal{C}^2$ on $\rp$.
Additionally, $\phi$ is the unique classical solution of the following HJB equation on $\rp:$
\beq\label{eq:2.5}
\begin{split}
&  \la \, f = (\tilde r z - 1) \, f' + {1 \over 2} \, b^2 \, (1 - \rho^2) \, z^2 \, f'' + \min_{\tilde \pi } \left[ (\mu - r - \sigma b \rho) \, \tilde \pi \, f' + {1 \over 2} \, \sigma^2 \, \tilde \pi^2 \, f'' \right], \\
&  f(0) = 1 \hbox{ and } \lim_{z \to \infty} f(z) = 0.
\end{split}
\endeq
The optimal investment strategy $\tilde \pi^*$ is given in feedback form by
\beq 
\tilde \pi^*_t = - {\mu - r - \sigma b \rho \over \sigma^2} \, {\phi'(Z^*_t) \over \phi''(Z^*_t)},
\endeq
in which $Z^*$ is the optimally controlled wealth process.
\end{thm}

\begin{proof} 
See Section~\ref{sec:3.1}, items 8 through 10, for an outline of the proof of this theorem and part of Theorem~\ref{thm:2.3}, and see Section~\ref{sec:3.3} for the proof itself. 
\end{proof}

The proof of the previous theorem under this assumption is performed by constructing the following sequence of functions, which we show (in Section~\ref{sec:3}) to be a regular sequence of functions that converges uniformly to $\phi$:
Consider the hitting time $\tilde \tau^z_M = \inf \{t \ge 0: Z_t \ge M \}$, for $M > 0$.  If we were to suppose that the goal of the individual were to minimize the probability of her wealth hitting 0 before dying {\it or} before
her wealth hitting $M > 0$, then we would have the modified minimum probability of lifetime ruin as follows:
\beq \label{eq:2.7}
\phi_M(z) = \inf_{\tilde \pi \in \widetilde{\mathcal{A}}}\widetilde{\P} \left( \tilde \tau^z_0 < (\tilde \tau^z_M \wedge \tau_d) \right),
\endeq
\noi Clearly, $\phi_M(z) = 0$ for $z \ge M$.  We have the following theorem for $\phi_M$.
\begin{thm}\label{thm:2.3}
The modified minimum probability of lifetime ruin $\phi_M$ given in $\eqref{eq:2.7}$ is continuous on $\rp$. Moreover, it is
strictly decreasing, strictly convex, and $\mathcal{C}^2$ on $(0, M)$.  Additionally, $\phi_M$ is the unique solution of the following HJB equation on $[0, M]:$
\beq \label{eq:2.8}
\begin{split}
&  \la \, f = (\tilde r z - 1) \, f' + {1 \over 2} \, b^2 \, (1 - \rho^2) \, z^2 \, f'' + \min_{\tilde \pi } \left[ (\mu - r - \sigma b \rho) \, \tilde \pi \, f' + {1 \over 2} \, \sigma^2 \, \tilde \pi^2 \, f'' \right], \\
&  f(0) = 1 \hbox{ and } f(M) = 0.
\end{split}
\endeq
The optimal investment strategy $\tilde \pi^*_M$ on $(0, M)$ is given in feedback form by
\beq \label{eq:2.9}
(\tilde \pi^*_M)_t = - {\mu - r - \sigma b \rho \over \sigma^2} \, {\phi'_M(Z^*_t) \over \phi''_M(Z^*_t)}\quad ,
\endeq
in which $Z^*$ is the optimally controlled wealth process. Furthermore, on $\rp,$ we have
\beq \label{eq:2.10}
\lim_{M \to \infty} \phi_M(z) = \phi(z). 
\endeq
In fact, the convergence is uniform.
\end{thm}
\begin{proof} See Section~\ref{sec:3.1}, items 6 through 10, for an outline of the proof of this theorem and part of Theorem~\ref{thm:2.2}, and see Section~\ref{sec:3.3} for the proof itself. 
\end{proof}
 In the proof of the regularity and the uniform convergence of $\{ \phi_M \}$ to $\phi$ as $M \to \infty$, we use the fact that each $\phi_M$ is the Legendre transform of a concave function, which is defined in the next section.  
\subsection{Controller-and-Stopper Problem}\label{sec:2.3}

For a positive constant $M$, define the following ``payoff function'' $u_M$ for $y \ge 0$ by
\beq \label{eq:2.11}
u_M(y) := \min(My, 1).
\endeq
\noi Fix values $y_M < 1/M < y_0$.  Note that $u$ is maximal among non-decreasing, concave functions $f$ defined on $\rp$ that take values $f(0)=0$, $f(y_M) = M y_M$ and $f(y_0) = 1$.

Define a controlled stochastic process $Y^{y,\a}$ by
\beq \label{eq:2.12}
dY^{y,\a}_t = Y^{y,\a}_t \left[ (\la - \tilde r) \, dt + {\mu - r - \sigma b \rho \over \sigma} d \hat B^{(1)}_t \right] + \a_t \left[ b
\sqrt{1 - \rho^2} \, dt + d \hat B^{(2)}_t  \right], 
\endeq
\noindent with $Y^{y,\a}_0 = y \geq 0$ in which $\hat B^{(1)}$ and $\hat B^{(2)}$ are independent standard Brownian motions on a probability space $(\hat{\Omega},\hat{\mathcal{F}},\hat{\P})$. We define the set of admissible controls as $\mathcal{A}(y)$ the collection of  ${\{ \hat{\mathcal{F}}_t \}}$-progressively measurable strategies $\alpha \equiv  (\a_t)_{t \geq 0}$ 
 (in which $\hat{\mathcal{F}}_t$ is the augmentation of $\sigma(\hat{B}^{(1)}_s, \hat{B}^2_s: 0 \le s \le t)$) that satisfy the integrability condition $\hat \E[\int_0^t \alpha_s^2 \, ds] < \infty$ and $Y^{y,\alpha}_t \geq 0$, almost surely, for all $t \ge 0$.
 
 Consider the controller-and-stopper problem given by
\beq \label{eq:2.13}
\hat \phi_M(y) = \sup_{\a \in \mathcal{A}(y)} \inf_\tau  \hat \E \left[ \int_0^{\tau} e^{-\la t} \, Y^{y,\a}_t \, dt + e^{-\la \tau} \,
u_M(Y^{y,\a}_\tau) \right],
\endeq
\noi in which $\tau$ is a stopping time with respect to $(\hat{\mathcal{F}}_t)_{t \geq 0}$.  For this problem, the controller receives a (discounted) running reward of $Y^{y,\a}$ and seeks to make this as high as possible until the stopper ends the game with the payoff of $u_M$.

 Although it is not obvious, it turns out that $\hat \phi_M$ is the concave dual of $\phi_M$, as we will show later in Section~\ref{sec:2.4}.  It is remarkable that the concave dual of the probability of ruin is the value function for a problem that has economic meaning.  Moreover, it is the value function for a class of games that has not been studied to a great extent, with the exception of the work of Karatzas and his co-authors (as referenced in the Introduction); therefore, analyzing this value function is of interest on its own right. We have the following theorem for the value function $\hat \phi_M$ defined in \eqref{eq:2.13}.

\begin{thm}\label{thm:2.4}
(i) The controller-and-stopper problem in \eqref{eq:2.13} has a continuation region given by $D = \{y \in \rp: \hat{\phi}_M(y)< u_M(y)\}=(y_M, y_0)$ for some $0 \le y_M \le 1/M \le y_0$.  The value function for this problem, namely $\hat \phi_M,$ is non-decreasing $($strictly increasing on $[0,y_0] \,
)$, concave, and $\mathcal{C}^2$ on $\rp$ $($except for possibly at $y_M$ and $y_0,$ where it is $\mathcal{C}^1)$. The value function is strictly concave on $(y_M,y_0)$.
(ii) Let us define $m$ by
\beq \label{eq:2.14}
m = {1 \over 2} \left( {\mu - r - \sigma b \rho  \over \sigma} \right)^2.
\endeq
Then $\hat \phi_M$ is the unique classical solution of the following boundary value problem among the positive functions bounded above by $u_M$:
\beq \label{eq:2.15}
\begin{split}
& \la g = y + (\la - \tilde r) y g' + m y^2 g'' + \max_{\a} \left[ b \sqrt{1 - \rho^2} \a g' + {1 \over 2} \a^2 g'' \right] \; \hbox{ on } \; D, \\
&g(y_M) = M y_M \hbox{ and } g(y_0) = 1.
\end{split}
\endeq
 Finally, $\hat \phi_M$ satisfies smooth pasting at the boundary of $D$; specifically, $\hat \phi_M'(y_M) = M$ and $\hat \phi_M'(y_0) = 0$.
\end{thm}

\begin{proof}See Section~\ref{sec:3.1}, items 1 through 5, for an outline of the proof of this theorem, and see Section~\ref{sec:3.2} for the proof itself.
\end{proof}

The optimal investment strategies for the controller and stopper problem are constructed Section 3 in Proposition~\ref{prop:3.9}.

\subsection{Relationship Among the Three Control Problems}\label{sec:2.4}

In this section, we show how the three control problems described in Sections \ref{sec:2.1} through \ref{sec:2.3} are related given the validity of Theorems \ref{thm:2.1} through \ref{thm:2.4}.  In Section~\ref{sec:3}, this relationship is heavily used to prove regularity of the problem of minimizing the lifetime probability of ruin.  A direct proof of regularity of this function is not available, whereas a direct proof of regularity of the controller-and-stopper problem is.

 Begin with $\hat \phi_M$, the value function for the controller-and-stopper problem defined in \eqref{eq:2.13}.  Because $\hat \phi_M$ is concave on $\rp$, we can define its convex dual via the Legendre transform (see, for example, Karatzas and Shreve (1998, Section 3.4))  as follows:
\beq \label{eq:2.18}
\Phi_M(z) = \max_{y \ge 0} \left[ \hat \phi_M(y) - zy \right].
\endeq

\begin{thm}\label{thm:2.5}
The convex dual, namely $\Phi_M$ of $\hat \phi_M$ is a $\mathcal{C}^2$ solution of the boundary-value problem $\eqref{eq:2.8}$ on $[0, M]$ with $\Phi_M(z) = 0$ for $z \ge M$. The function $z \to \Phi_M(z)$ is strictly decreasing on $[0,M]$ and strictly convex on $z \in (0,M)$.
\end{thm}
\begin{proof} We have two cases to consider:  (1) $z \ge M$ and (2) $z < M$.  If $z \ge M$, then $\Phi_M(z) = 0$ because $\hat \phi_M(y) \le u_M(y) \le M y \le z y$, from which it follows that the maximum on the right-hand side of \eqref{eq:2.18} is achieved at $y^* = y_M$.

For the remainder of this proof, assume that $z < M$. In this case, the critical value $y^*$ solves the equation $\hat \phi'_M(y) - z = 0$.  (Recall that the slope of $\hat \phi_M$ decreases from $M$ at $y = y_M$ to 0 at $y = y_0$ continuously.)  Thus, $y^* = I_M(z)$, in which $I_M$ is the inverse of $\hat \phi'_M$ on $[y_M,  y_0]$.  It follows that for $z < M$, we have
\beq \label{eq:2.19}
\Phi_M(z) = \hat \phi_M \left[ I_M(z) \right] - z I_M(z). 
\endeq

\noi Expression \eqref{eq:2.19} implies that
\beq \label{eq:2.20}
\Phi'_M(z) = \hat \phi'_M \left[ I_M(z) \right] I'_M(z) - I_M(z) - z I'_M(z) = z I'_M(z) - I_M(z) - z I'_M(z)
= - I_M(z)<0, \, z\in [0,M].
\endeq
\noi Thus, the dual variable $y$ is related to the original variable $z$ via $y^* = I_M(z) = - \Phi'_M(z)$. Note that from \eqref{eq:2.20}, we have
$$\Phi''_M(z) = -I'_M(z) = -1/\hat \phi''_M \left[ I_M(z) \right].$$
Note that $z \to \Phi_M(z)$ is strictly convex on $(0,M)$, since $y \to \hat{\phi}_M(y)$ is strictly concave on $(y_M,y_0)$.
We proceed to find the boundary-value problem that $\Phi_M$ solves given that $\hat \phi_M$ solves the free-boundary problem in \eqref{eq:2.15}. In the differential equation for $\hat \phi_M$ in \eqref{eq:2.15} with the optimal control substituted for $\a$, let $y = I_M(z) = -\Phi'_M(z)$ to obtain
$$\la \hat{\phi}_M \left[ I_M(z) \right] = I_M(z) + (\la - \tilde r) I_M(z) \hat \phi'_M \left[ I_M(z) \right] + m I^2_M(z)
\hat \phi''_M \left[ I_M(z) \right] 
- {1 \over 2} b^2 (1 - \rho^2) { \left( \hat \phi'_M \left[ I_M(z) \right] \right)^2 \over \hat \phi''_M \left[ I_M(z) \right]}.$$
\noi Rewrite this equation in terms of $\Phi_M$ to get
$$\la \left[ \Phi_M(z) - z \Phi'_M(z) \right] = - \Phi'_M(z) - (\la - \tilde r) z \Phi'_M(z) + m {(\Phi'_M(z))^2 \over -
\Phi''_M(z)}  - {1 \over 2} b^2 (1 - \rho^2) {z^2 \over -1/\Phi''_M(z)},$$
\noi or equivalently,
\beq \label{eq:2.21}
\la \Phi_M(z) = (\tilde r z - 1) \Phi'_M(z) - m{(\Phi'_M(z))^2 \over \Phi''_M(z)} + {1 \over 2} b^2 (1 - \rho^2) z^2 \Phi''_M(z),
\endeq
\noi which is identical to the differential equation in \eqref{eq:2.8} that $\phi_M$ solves on $[0, M]$.

Next, consider the boundary conditions in \eqref{eq:2.15}.  The boundary conditions at $y = y_M$, namely $\hat \phi_M(y_M) = M y_M$ and $\hat \phi'_M(y_M) = M$, imply that the corresponding dual value of $z$ is $M$ and that
\beq \label{eq:2.22}
\Phi_M(M) = 0, 
\endeq
\noi as earlier discussed in the case when $z \ge M$.  Similarly, the boundary conditions at $y = y_0$, namely $\hat \phi_M(y_0) = 1$ and $\hat \phi'_M(y_0) = 0$, imply that the corresponding dual value of $z$ is 0 and that
\beq \label{eq:2.23}
\Phi_M(0) = 1. 
\endeq
Thus, we have shown that the Legendre transform $\Phi_M$ of the value function of the optimal controller-and-stopper problem $\hat \phi_M$ in \eqref{eq:2.13}, or equivalently the solution of the free-boundary problem in \eqref{eq:2.15}, is the solution of the boundary-value problem \eqref{eq:2.21}-\eqref{eq:2.23} on $[0, M]$.  Note that the boundary-value problem for $\Phi_M$ is {\it identical} to the one for $\phi_M$ in \eqref{eq:2.8}.  Additionally, we showed that for $z \ge M$, $\Phi_M(z) = 0$, which is also clearly true for $\phi_M$.  \end{proof}

In other words, we have shown that under the validity of Theorems \ref{thm:2.3} and \ref{thm:2.4}, the Legendre transform of $\hat \phi_M$ equals the minimum probability of ruin $\phi_M$.  Next, note that it is natural that $\lim_{M \to \infty} \phi_M(z) = \phi(z)$ on $\rp$, and we show this result below in Section~\ref{sec:3.3}; see Propositions \ref{prop:3.14} through \ref{prop:3.16}.

Finally, we relate $\phi$ in \eqref{eq:2.4} and \eqref{eq:2.5} to $\psi$ in \eqref{eq:2-1} and \eqref{eq:2.2}.  Indeed, define $\Psi$ on $\rp^2$ by $\Psi(w, c) = \phi(w/c)$.  Then, after a fair amount of calculus, one can show that $\Psi$ solves \eqref{eq:2.2}. By the uniqueness of the solution of \eqref{eq:2.2}, it follows that $\Psi = \psi$.  In other words, $\phi$ and $\psi$ are related by $\phi(z) = \psi(z, 1)$ and $\psi(w, c) = \phi(w/c)$.  Moreover, the optimal investment strategy $\pi^*$ for the problem in Section~\ref{sec:2.1} is related to the optimal investment strategy $\tilde \pi^*$ for the problem in Section~\ref{sec:2.2} by
\beq \label{eq:2.24}
\pi^*_t = c \left(\tilde \pi^*_t + \rho \, {b \over \sigma} \, Z^*_t \right),
\endeq
\noi in which $Z^*$ is the optimally-controlled wealth.
\section{Proofs of Theorems 2.1-2.4}\label{sec:3}

In this section, we prove Theorems \ref{thm:2.1} through \ref{thm:2.4}, namely, that each of the value functions for the problems described in Sections~\ref{sec:2.1} through \ref{sec:2.3} is smooth and is the unique solution of its respective HJB.  We will see that Theorems \ref{thm:2.4} and \ref{thm:2.5} are the primary ingredients of the proof of our main theorem, Theorem~\ref{thm:2.1}.

In Section~\ref{sec:3.1}, we outline our program for proving these theorems.  In Section~\ref{sec:3.2}, we prove Theorem~\ref{thm:2.4} via a series of propositions.  In Section~\ref{sec:3.3}, we first prove Theorem \ref{thm:2.3} using Theorem~\ref{thm:2.4}. Theorem~\ref{thm:2.2} follows as a corollary of Theorem~\ref{thm:2.3}.  Finally, we prove Theorem~\ref{thm:2.1} in Section~\ref{sec:3.4} using Theorem~\ref{thm:2.2}. 
\subsection{Scheme for Proving Theorems~\ref{thm:2.1}-\ref{thm:2.4}}\label{sec:3.1}

To show that value functions for the three problems have the properties stated in Theorems~\ref{thm:2.1}-\ref{thm:2.4} and to show that each is the unique solution of its corresponding HJB, we will proceed as follows:
\medskip

\noindent {\it Proof of Theorem~\ref{thm:2.4} and extensive analysis of the controller-and-stopper problem:}

\smallskip

\begin{itemize}
\item[1.] Show that $\hat \phi_M$ is a viscosity solution of the HJB-VI \eqref{eq:2.15}; see Propositions \ref{prop:3.1} through \ref{prop:3.3}.

\item[2.] Prove a comparison theorem for \eqref{eq:2.15}; see Proposition~\ref{prop:3.6}.  From this result, conclude that $\hat \phi_M$ is the {\it unique} viscosity solution of the HJB-VI; see Corollary~\ref{cor:3.7}.

\item[3.] Show that smooth pasting holds for the controller-and-stopper problem; see Proposition~\ref{prop:3.10}.

\item[4.] Show that $\hat \phi_M$ is $\mathcal{C}^2$ and strictly concave in the continuation region; see Propositions \ref{prop:3.8}. From this result, conclude that $\hat \phi_M$ is in $\mathcal{C}^1(\rp) \cap \mathcal{C}^2(\rp - \{y_M\} - \{y_0\})$, in which $(y_M, y_0)$ is the continuation region.

\item[5.] Conclude that $\hat \phi_M$ is the unique solution in $\mathcal{C}^1(\rp) \cap \mathcal{C}^2(\rp - \{y_M\} - \{y_0\})$ of the free-boundary problem on $\rp$ given in \eqref{eq:2.15}; see Corollary
~\ref{cor:3.11} and Proposition~\ref{prop:3.9}. The latter also constructs optimal stopping and control strategies for the controller and stopper problem.

\medskip

\noindent {\it Proof of the regularity portion of Theorem~\ref{thm:2.3} as a corollary of Theorem~\ref{thm:2.4}:}

\smallskip

\item[6.] Then, from Theorem~\ref{thm:2.5} in Section~\ref{sec:2.4}, conclude that the convex dual, namely $\Phi_M$, of $\hat \phi_M$ (via the Legendre transform) is a $\mathcal{C}^2$ solution of \eqref{eq:2.8} on $[0, M]$ with $\Phi_M(z) = 0$ for $z \ge M$.

\item[7.] Show via a verification lemma that the minimum probability of ruin $\phi_M$ defined in \eqref{eq:2.7} equals $\Phi_M$; see Lemma~\ref{lem:3.12} and Proposition~\ref{prop:3.13}.

\medskip

\noindent {\it Proof of the limit result of Theorem~\ref{thm:2.3} and of Theorem~\ref{thm:2.2} as a corollary of Theorem~\ref{thm:2.3}:}

\smallskip

\item[8.] Show that $\lim_{M \to \infty} \phi_M$ is a viscosity solution of \eqref{eq:2.5}; see Proposition~\ref{prop:3.14}.

\item[9.] Show that $\lim_{M \to \infty} \phi_M$ is smooth; see Proposition~\ref{prop:3.15}.

\item[10.] Show that $\lim_{M \to \infty} \phi_M = \phi$ on $\rp$ and that $\phi$ is the unique smooth solution of \eqref{eq:2.5}; see Proposition~\ref{prop:3.16}.

\medskip

\noindent {\it Proof of Theorem~\ref{thm:2.1} as a corollary of Theorem~\ref{thm:2.2}:}

\smallskip

\item[11.]  Because $\phi$ is a classical solution of \eqref{eq:2.5}, it follows that $(w, c) \to \phi(w/c)$ defines a classical solution of \eqref{eq:2.2}.  Then, via a verification lemma, we conclude that the minimum probability of ruin $\psi$ defined in \eqref{eq:2-1} is given by $\psi(w, c) = \phi(w/c)$; see Proposition~\ref{prop:3.17}.

\end{itemize}

In other words, our primary goal is to show that the minimal probability of ruin for the problem in Section~\ref{sec:2.1} is the unique classical solution of its HJB equation; that is, show that it is {\it regular}.  It is not clear {\it a priori} that the value functions of the two problems in Section \ref{sec:2.1} and \ref{sec:2.2} are convex or smooth, and we prove their regularity via the related problem in Section \ref{sec:2.3}.  Specifically, we (a) show that $\hat \phi_M$ is regular (items 1 through 5 above); (b) then, we show that the convex dual of $\hat \phi_M$ equals $\phi_M$ and is regular (items 6 and 7 above); (c) then, we show that $\lim_{M \to \infty} \phi_M = \phi$ and is regular (items 8 through 10 above) by using the fact that $\phi$ is uniformly approximated by a regular sequence of functions; and (d) finally, we show that $\psi(w, c) = \phi(w/c)$ and that $\psi$ is regular.

\subsection{Proof of Theorem~\ref{thm:2.4} and an Extensive Analysis of the Controller-and-Stopper Problem}\label{sec:3.2}

To prove Theorem~\ref{thm:2.4}, we begin with a series of propositions that give us useful properties of $\hat \phi_M$.

\begin{prop} \label{prop:3.1}
$\hat \phi_M \le u_M$ on $\rp$, and $\hat \phi_M(0) = 0$.  Moreover, $\hat \phi_M$ is non-decreasing, concave, and continuous on $[0,\infty)$. In fact, $\hat \phi_M$ is uniformly continuous on $[0,\infty)$.
\end{prop}

\pf  That $\hat \phi_M \le u_M$ on $\rp$ is clear.  If $y = 0$, then because $\hat \phi_M(0) \le u_M(0) = 0$, the best that the
stopper can do is to set $\tau=0$ so that $\hat \phi_M(0)=u_M(0)=0$.

For $y > 0$ and $\a \in \mathcal{A}(y)$, it is straightforward to show that if $Y^{y,\a}_0 = y$, then

\beq \label{eq:3.1}
Y^{y,\a}_t = H_t \left( y + \int_0^t \frac{\a_s}{H_s} \left[ b \sqrt{1 - \rho^2} ds + d \hat B^{(2)}_s \right] \, \right), 
\endeq
\noi in which $H$ is the process defined by

$$H_t = \exp \left( \left( \lambda - \tilde r - {1 \over 2} \left( {\mu - r - \sigma b \rho \over \sigma} \right)^2 \right) t + {\mu
- r - \sigma b \rho \over \sigma} \hat B^{(1)}_t \right).$$

Now, suppose $0 < y_1 < y_2$; then, for $\a \in \mathcal{A}(y_1)$, we have $\a \in \mathcal{A}(y_2)$.  From \eqref{eq:3.1}, it follows that $Y^{ y_1,\a} < Y^{y_2,\a}$. Then, because the expression in the expectation of \eqref{eq:2.13} is non-decreasing with respect to $Y^{y,\a}$, we conclude that $\hat \phi_M$ is non-decreasing on $\rp$.

Next, we prove that the function $\hat \phi_M$ is concave:
\begin{equation*}
\begin{split}
&\hat{\phi}_{M}(\omega y_1 + (1 - \omega) y_2)=\sup_{\a \in  \mathcal{A}(\omega y_1 + (1 - \omega) y_2)} \inf_{\tau} \hat \E \left[ \int_0^{\tau} e^{-\la t} \, Y^{\omega y_1 + (1 - \omega) y_2, \a}_t \, dt + e^{-\la \tau} \,
u_M(Y^{\omega y_1 + (1 - \omega) y_2, \a}_\tau) \right]
\\& \geq \sup_{\alpha_1 \in  \mathcal{A}(y_1), \alpha_2 \in  \mathcal{A}(y_2)} \inf_{\tau}
\hat \E \bigg[ \int_0^{\tau} e^{-\la t} \, Y^{\omega y_1 + (1 - \omega) y_2, \omega \a_1+ (1-\omega)\a_2}_t \, dt  +e^{-\la \tau} \,
u_M\left(Y^{\omega y_1 + (1 - \omega) y_2, \omega \a_1+ (1-\omega)\a_2}_{\tau}\right) \bigg]
\\& \geq \sup_{\alpha_1 \in  \mathcal{A}(y_1), \alpha_2 \in  \mathcal{A}(y_2)} \inf_{\tau}
\hat \E \left[ \int_0^{\tau} e^{-\la t} \, \left(\omega Y^{ y_1,\a_1} + (1 - \omega) Y^{y_2, \a_2}\right) \, dt + e^{-\la \tau} \, \left( \omega
u_M(Y^{y_1,\a_1}_\tau)+(1-\omega)u_M(Y^{y_2,\a_2}_\tau)\right) \right] \\&
\geq \omega \hat{\phi}_{M}(y_1)+(1-\omega) \hat{\phi}_{M}(y_2).
\end{split}
\end{equation*}
Here, the first inequality follows since for any $\a_i \in \mathcal{A}(y_i)$ for $i = 1, 2$,  we have that $\omega \a_1 + (1 - \omega) \a_2 \in \mathcal{A}(\omega y_1 + (1 - \omega) y_2)$. The second inequality follows since $Y^{y,a}$ is a linear in both $y$ and $\alpha$ (see \eqref{eq:3.1}), and $u_M$ is concave.

Because $\hat \phi_M$ is concave on $\rp$, the only place that it might be discontinuous is at $y = 0$.  However, $\hat \phi_M(0) = 0$ and $\hat \phi_M \le u_M$, so $\hat \phi_M$ does not have a discontinuity at $y = 0$. Therefore, we conclude that $\hat \phi_M$ is continuous on $\rp$.

Because $\hat \phi_M$ is non-decreasing, concave, and is dominated by $u_M$ (which implies that the slope at 0 is bounded by $M$), it follows that
\beq \label{eq:uni-lips}
|\hat \phi_M(y) -\hat \phi_M(x)| \leq M |y-x|,
\endeq
\noi for any $(x,y) \in \rp^2$.  This Lipschitz continuity of $\hat \phi_M$ implies that it is uniformly continuous on $[0,\infty)$. \end{proof}

Define the region
\beq \label{eq:3.2}
D = \{ y \in \rp: \hat \phi_M(y) < u_M(y) \}, 
\endeq

\noi Later, in Proposition~\ref{prop:3.9}, we show that $D$ is the \emph{continuation region} for this controller-and-stopper problem.  That is, it is optimal for the stopper to let the game continue if and only if $y \in D$.

\begin{prop} \label{prop:3.2} There exist $0 \le y_M \le 1/M \le y_0 \le \infty$ such that $D = (y_M, y_0)$. \end{prop}

\begin{proof} Suppose that $y_1 > 0$ is such that $\hat \phi_M(y_1) = u_M(y_1)$.  First, suppose that $y_1 \le 1/M$; then, because $\hat \phi_M(0) = 0$ and because $\hat \phi_M$ is non-decreasing, concave, and bounded above by the line $My$ it must be that $\hat \phi_M(y) = My$ for all $0 \le y \le y_1$.  Thus, if $y_1 \le 1/M$ is not in $D$, then the same is true for $y
\in [0, y_1]$.

Finally, suppose that $y_1 \ge 1/M$; then, because $\hat \phi_M$ is non-decreasing, concave, and bounded above by the horizontal line $1$ it must be that $\hat \phi_M(y) = 1$ for all $y \ge y_1$.  Thus, if $y_1 \ge 1/M$ is not in $D$, then the
same is true for $y \in [y_1, \infty)$.

It follows that there exist $0 \le y_M \le 1/M \le y_0 \le \infty$ such that $D = (y_M, y_0)$.  Note that if $D$ is empty, we can
take $y_M = 1/M = y_0$. \end{proof}

We want to show that $\hat \phi_M$ is the unique solution of \eqref{eq:2.15} and that it is $\mathcal{C}^2$, except possibly at $y_M$ and $y_0$.  To this end, we first show that $\hat \phi_M$ is a viscosity solution, in which we define a viscosity solution as follows:

{\defi  \label{def:3.1} $(i)$ $g \in \mathcal{C}(\R_+)$ is a {\rm viscosity supersolution (}respectively, {\rm subsolution)} of the controller-stopper problem if for all $y_1 \in \R_+$ it holds that
\beq \label{eq:3.3}
\begin{split}
\max & \bigg[ \la g(y_1) - y_1 - (\la - \tilde r) y_1 f'(y_1) - m y_1^2 f''(y_1) - \max_{\a} \left[ b \sqrt{1 - \rho^2} \,
\a f'(y_1) + {1 \over 2} \a^2 f''(y_1) \right], \\
& \quad g(y_1) - u_M(y_1) \bigg] \ge 0,
\end{split}
\endeq
$($respectively, $\le 0)$ whenever $f \in  \mathcal{C}^2(\R_+)$ and $g - f$ has a global minimum $($respectively, maximum$)$. \hfill \break
\noi$(ii)$ $g$ is a {\rm viscosity solution} if it is both a viscosity super- and subsolution.}

We will use the dynamic programming principle for differential games, see Fleming and Souganidis (1989). Although this theorem is stated in terms of two controllers we can still apply this theorem by turning the stopper into a controller by assigning $v_{\tau}=1_{\{\tau<t\}}$ for each stopping time $\tau$.
As a result of the dynamic programming principle (DPP), $\hat \phi_M$ is a viscosity solution of the HJB-VI in \eqref{eq:3.3}, as we
show in the following proposition:

{\prop \label{prop:3.3} The function $\hat \phi_M$ is a viscosity solution of the HJB-VI in \eqref{eq:3.3}.}

\pf  The proof follows from Theorem 4.1 of Buckdahn and Li (2009). Here, we will present the proof of the viscosity subsolution property using more classical arguments similar to the ones used in Proposition 4.3.2 in Pham (2009). 

First recall that $\hat \phi_M \in  \mathcal{C}(\rp)$ from Proposition~\ref{prop:3.1}.  We will show that 
\beq \label{eq:w_L}
\begin{split}
\max & \bigg[ \la f(y_1) - y_1 - (\la - \tilde r) y_1 f'(y_1) - m y_1^2 f''(y_1) - \max_{\a} \left[ b \sqrt{1 - \rho^2} \,
\a f'(y_1) + {1 \over 2} \a^2 f''(y_1) \right], \\
& \quad f(y_1) - u_M(y_1) \bigg] \leq 0,
\end{split}
\endeq
for $f \in \mathcal{C}^2(\R_+)$ such that $\hat \phi_M - f$ has a global maximum at $ y_1$.  Without loss of generality, we can assume that $\hat \phi_M(y_1) = f(y_1)$, $\hat \phi_M \le f$ on $\R_+$, and that $y_1$ is a strict maximum point of $\hat \phi_M-f$.  Because $\hat \phi_M(y_1) \le u_M(y_1)$, it is enough to prove the following inequality at $ y_1$:
\beq \label{eq:3.4}
\la f(y_1) \le y_1 + (\la - \tilde r) y_1 f'(y_1) + m y_1^2 f''(y_1) + \max_{\a} \left[ b \sqrt{1 - \rho^2} \, \a f'(y_1)
+ {1 \over 2} \a^2 f''(y_1) \right].  
\endeq
We will prove that \eqref{eq:3.4} holds by contradiction. Assume that
\begin{equation} \label{eq:saty1}
\la f(y_1) > y_1 + (\la - \tilde r) y_1 f'(y_1) + m y_1^2 f''(y_1) + \max_{\a} \left[ b \sqrt{1 - \rho^2} \, \a f'(y_1)
+ {1 \over 2} \a^2 f''(y_1) \right]
\end{equation}
Observe that \eqref{eq:saty1} holds for a test function $f$ if either $f''(y_1)<0$ or $f''(y_1)=f'(y_1)=0$. First, we will show that the latter scenario is not possible. To this end, observe that
\begin{equation}\label{eq:rderophi}
0=\lim_{h \downarrow 0} \frac{f(y_1+h)-f(y_1)}{h} \geq \lim_{h \downarrow 0} \frac{\hat{\phi}_M(y_1+h)-\hat{\phi}_M(y_1)}{h}:=D_{+}\hat{\phi}_M(y_1),
\end{equation}
where the last expression denotes the right-derivative of $\hat{\phi}_M$, which always exits thanks to the concavity of $\hat{\phi}_M$. Since $\hat{\phi}_M$ is increasing, \eqref{eq:rderophi} implies that $D_{+}\hat{\phi}_M(y_1)=0$. Now, using the fact that $\hat{\phi}_M$ is increasing again, we obtain that $\hat{\phi}_M'(y)=0$ for $y>y_1$. We can conclude that $\hat{\phi}_M(y)=\hat{\phi}_M(y_1)$, $y \geq y_1$. Let us define $\tau^{\eps}(\a):=\inf\{t \geq 0: Y_{t}^{y_1,\a} \geq y_1+\eps\}$,  with the convention that $\inf \emptyset =\infty$. 
The DPP applied to \eqref{eq:2.13} yields
\begin{equation}\label{eq:wwgac}
\begin{split}
\hat{\phi}_M(y_1) &= \sup_{\a \in \mathcal{A}(y_1)} \inf_\tau  \hat \E \left[ \int_0^{\tau \wedge \tau^{\eps}(\a)} e^{-\la t} \, Y^{y_1,\a}_t \, dt +  e^{-\la (\tau \wedge \tau^{\eps}(\a))} \, \hat{\phi}_M \left(Y^{y_1,\a}_{\tau^{\eps}(\a) \wedge \tau}\right) \right]
\\ &\leq \sup_{\a \in \mathcal{A}(y_1)}\inf_\tau  \hat \E \left[ \int_0^{\tau \wedge \tau^{\eps}(\a)} e^{-\la t} (y_1+\eps) dt +  e^{-\la (\tau \wedge \tau^{\eps}(\a))} \,\hat{\phi}_M(y_1) \right]
\\& = \frac{y_1+\eps}{\la}+\sup_{\a \in \mathcal{A}(y_1)} \inf_\tau   \hat \E \left[\left(\hat{\phi}_M(y_1)-\frac{(y_1+\eps)}{\la}\right) e^{-\la (\tau \wedge \tau^{\eps}(\a))}\right].
\end{split}
\end{equation}
We will show that \eqref{eq:wwgac} implies 
\begin{equation}\label{eq:tphimineq}
\hat{\phi}_M(y_1) \leq \frac{y_1+\eps}{\la}.
\end{equation}
Let us assume the contrary. Then \eqref{eq:wwgac} implies that
\begin{equation}\label{eq:wgacttt}
\sup_{\a \in \mathcal{A}(y_1)}\hat{\E}\left[e^{-\la \tau^{\eps}(\a)}\right] = 1.
\end{equation}
Let us define 
\[
u(y):= \sup_{\a \in \mathcal{A}(y)}\hat{\E}\left[e^{-\la \tilde{\tau}^{\eps}(\a)}\right], \quad \text{where} \quad \tilde{\tau}^{\eps}(\a):=\inf\{t \geq 0: Y_{t}^{y,\a} \geq y_1+\eps\},
\]
which is an example of a stochastic exit time problem. It follows that $u$ is a viscosity solution of 
\begin{equation}\label{eq:setcp}
\lambda u(y)- (\lambda-\tilde{r})y u'(y)-my^2 u''(y)-\max_{\a} \left[ b \sqrt{1 - \rho^2} \, \a u'(y_1)
+ {1 \over 2} \a^2 u''(y_1) \right]=0, \quad y \in (0,y_1+\eps),
\end{equation}
with boundary condition $u(y_1+\eps)=1$; see e.g. Bayraktar et al. (2010). Since $u(y_1)=1$ (see \eqref{eq:wgacttt}), and $u \leq \tilde{u} \equiv 1$, it follows from the viscosity subsolution property of $u$ (applying the subsolution inequality to the test function $\tilde{u} \equiv 1$) that 
$\lambda \leq 0$, which contradicts the choice of $\lambda$. Therefore, $u(y_1)<1$ and as a result \eqref{eq:tphimineq} holds. Since $\eps$ is arbitrary we have that $\hat{\phi}_M(y_1) \leq y_{1}/\la$. However, this inequality contradicts \eqref{eq:saty1}, because this equation together with $f'(y_1)=f''(y_1)=0$ implies that $\hat{\phi}_M(y_1)=f(y_1)>y_1/\la$. Therefore, we can not have that $f'(y_1)=f''(y_1)=0$.

In what follows, we will assume, for the ease of notation, that $y_1>0$. Similar arguments to those below hold when $y_1=0$.
We have so far proven that \eqref{eq:saty1} (together with the fact that $\hat{\phi}$ is nondecreasing and concave) implies that $f''(y_1)<0$. As a result, for small enough $\delta>0$, there exists $\eta>0$ and $\eps \in (0, \delta \lambda]$ such that
\[
\max\{\hat{\phi}_M(y_1-\eta)-f(y_1-\eta), \hat{\phi}_M(y_1+\eta)-f(y_1+\eta)\} = -\delta,
\]
\beq\label{eq:cnthyp}
\la f(y) - y - (\la - \tilde r) y f'(y) - m y_1^2 f''(y) - \max_{\a} \left[ b \sqrt{1 - \rho^2} \, \a f'(y)
+ {1 \over 2} \a^2 f''(y) \right] \geq \eps,
\endeq
for $y \in (y_1-\eta, y_1+\eta)$. Given $T>0$ and $\a \in \mathcal{A}(y_1)$, define
\beq \label{eq:3.5}
 \theta(\alpha):=\inf \left\{t \ge 0: |Y^{y_1,\a}_t-y_1| \geq \eta \right\}, \quad \tau^{T}(\a):=T \wedge \theta(\a).
 \endeq
 with the convention that $\inf \emptyset=\infty$. The DPP applied to \eqref{eq:2.13} yields
\beq \label{eq:DPPfG}
\hat{\phi}_M(y_1) = \sup_{\a \in \mathcal{A}(y_1)} \inf_\tau  \hat \E \left[ \int_0^{\tau \wedge \tau^T(\a)} e^{-\la t} \, Y^{y_1,\a}_t \, dt +  e^{-\la (\tau \wedge \tau^T(\a))} \, \hat{\phi}_M \left(Y^{y_1,\a}_{\tau^T(\a) \wedge \tau}\right) \right],
\endeq
Let $\a^* \in \mathcal{A}(y_1)$ be an $\eps/(2 \lambda)$-optimal strategy for the right-hand-side of \eqref{eq:DPPfG}. Then
\[
\begin{split}
\hat{\phi}_M(y_1)-\frac{\eps}{2\lambda} &\leq  \inf_\tau  \hat \E \left[ \int_0^{\tau \wedge \tau^T(\a^*)} e^{-\la t} \, Y^{y_1,\a^*}_t \, dt +  e^{-\la (\tau \wedge \tau^T(\a^*))} \, \hat{\phi}_M\left(Y^{y_1,\a^*}_{\tau^T(\a^*) \wedge \tau}\right)\right]
\\&\leq  \hat \E \left[ \int_0^{ \tau^T(\a^*)} e^{-\la t} \, Y^{y_1,\a^*}_t \, dt +  e^{-\la \tau^T(\a^*)} \, \hat{\phi}_M \left(Y^{y_1,\a^*}_{\tau^T(\a^*)}\right)\right].
\end{split}
\]
\noi It follows that
\begin{equation}\label{eq:3.9becomes}
-\frac{\eps}{2\lambda}  \le \hat \E \left[ \int_0^{\tau^T(\a^{*})} e^{-\la t} \, Y^{y_1,\a^{*}}_t \, dt + e^{-\la {\tau^T(\a^{*})}} \,
f\left(Y^{y_1,\a^{*}}_{\tau^T(\a^{*})}\right) - f(y_1) - \delta e^{-\la \tau^T(\a^{*})}1_{\{\theta(\a^*) \leq T\}}\right].
\end{equation}
\noi By applying It\^o's formula to  $e^{-\la t} \, f\left(Y_t^{y_1,\a^{*}}\right)$, we obtain
\beq
\begin{split}
 & e^{-\la \tau^T(\a^{*})}f\left(Y^{y_1,\a^{*}}_{\tau^T(\a^{*})}\right) =   f(y_1) + \int_0^{\tau^T(\a^{*})} e^{-\la t} \left(-\la f\left(Y_t^{y_1,\a^{*}}\right) + \left((\la - \tilde r)Y_t^{y_1,\a^{*}} + \a^{*}_t b\sqrt{1 - \rho^2} \right) f'\left(Y_t^{y_1,\a^{*}}\right) \right) \, dt  \cr
& + \int_0^{\tau^T(\a^{*})} e^{-\la t} \left(m \cdot \left(Y_t^{y_1, \a^{*}}\right)^2 + {1 \over 2} (\a^{*}_t)^2 \right) f''\left(Y^{y_1,\a^{*}}_t\right) \, dt + {\mu - r -\sigma b \rho \over \sigma} \int_0^{\tau^T(\a^{*})} e^{-\lambda t} f'\left(Y_t^{y_1,\a^{*}}\right)Y_t^{y_1,\a^{*}} d \hat{B}_t^{(1)} \cr &+ \int_0^{\tau^T(\a^{*})} e^{-\lambda t} \a^{*}_t f'\left(Y^{y_1,\a^{*}}_t\right) d\hat{B}_t^{(2)},
\end{split}
\endeq
\noi which, thanks to the definition of $\tau^{T}(\a^{*})$ and to the fact that $\hat{\E}\left[\int_0^T (\a^*_s)^2 ds\right]<\infty$ (see the definition of admissible strategies), leads to
\beq \label{eq:3.7}
\begin{split}
 & \hat \E \left[e^{-\la {\tau^T(\a^{*})}} \, f\left(Y^{y_1,\a^*}_{\tau^T(\a^{*})}\right) - f(y_1)\right] \\ & = \hat \E\bigg[\int_0^{\tau^T(\a^{*})} e^{-\la t}\left(-\la f\left(Y_t^{y_1,\a^{*}}\right) + \left((\la - \tilde{r}) Y_t^{y_1,\a^{*}} + \a_t^{*} b \sqrt{1 - \rho^2} \right) f'\left(Y_t^{y_1,\a^{*}}\right) \right) \, dt \\
&\quad \; + \int_0^{\tau^T(\a^{*})} e^{-\la t} \left(m\cdot(Y_t^{y_1,\a^{*}})^2 + {1 \over 2} \left(\a^{*}_t\right)^2\right) f''\left(Y^{y_1,\a^{*}}_t\right) \, dt \bigg].
\end{split}
\endeq

Using \eqref{eq:cnthyp}, \eqref{eq:3.9becomes}, and \eqref{eq:3.7}, we can write
\begin{equation}\label{eq:blhtgz}
\begin{split}
-\frac{\eps}{2 \lambda}\le& \;  \, \hat \E \left[ \int_0^{\tau^T(\a^{*})} e^{-\la t}\left( Y_t^{y_1,\a^{*}} - \la f\left(Y_t^{y_1,\a^{*}}\right) + \left(\la- \tilde r\right) Y_t^{y_1,\a^{*}}
f'\left(Y_t^{y_1,\a^{*}}\right) + m \cdot \left(Y_{t}^{y_1,\a^{*}}\right)^2 f''\left(Y_{t}^{y_1,\a^{*}}\right) \right) dt \right] \cr &+ \, \hat \E \left[ \int_0^{\tau^T(\a^{*})}
e^{-\la t}\left( b \sqrt{1 - \rho^2} \, \a^{*}_t f'\left(Y_{t}^{y_1,\a^{*}}\right) + {1 \over 2} \left(\a^{*}_t\right)^2 f''\left(Y_{t}^{y_1,\a^{*}}\right) \right) dt - \delta e^{-\la \tau^T(\a^{*})}1_{\{\theta(\a^*) \leq T\}}\right] \\
\le &  \; \, \hat \E \left[ \int_0^{\tau^T(\a^{*})} e^{-\la t}\left( Y_t^{y_1,\a^{*}} - \la f\left(Y_t^{y_1,\a^{*}}\right) + \left(\la- \tilde r\right) Y_t^{y_1,\a^{*}}
f'\left(Y_t^{y_1,\a^{*}}\right) + m \cdot \left(Y_{t}^{y_1,\a^{*}}\right)^2 f''\left(Y_{t}^{y_1,\a^{*}}\right) \right) dt \right] \\
&+ \, \hat \E \left[ \int_0^{\tau^T(\a^{*})}
e^{-\la t}\max_{c}\left( b \sqrt{1 - \rho^2} \, c f'\left(Y_{t}^{y_1,\a^{*}}\right) + {1 \over 2} c^2 f''\left(Y_{t}^{y_1,\a^{*}}\right) \right) dt - \delta e^{-\la \tau^T(\a^{*})}1_{\{\theta(\a^{*}) \leq T\}}\right]
\\&\leq - \hat \E\left[\int_0^{\tau^T(\a^{*})}\eps e^{-\lambda t} dt+\delta e^{-\la \tau^T(\a^{*})}1_{\{\theta(\a^{*}) \leq T\}}\right]
\\&=-\frac{\eps}{\la}+\hat \E \left[\left(\frac{\eps}{\lambda}-\delta\right)e^{-\lambda \tau^T(\a^{*})}1_{\{\theta(\a^{*}) \leq T\}}+\frac{\eps}{\lambda}e^{-\lambda \tau^T(\a^{*})}1_{\{\theta(\a^{*})>T\}}\right]
\\& \leq -\frac{\eps}{\la}+ \hat \E \left[ \frac{\eps}{\lambda}1_{\{\theta(\a^{*})>T\}} e^{-\lambda \tau^T(\a^{*})} \right].
\end{split}
\end{equation}
We get a contradiction by sending $T$ to infinity: $-\eps/ (2 \la) \leq -\eps/\la$. 
\end{proof}

{\cor \label{cor:3.4} If $M > 1/\la$, then $D = (y_M, y_0)$ is non-empty.  In particular, $y_M < 1/M < \lambda \leq y_0$.}

\pf Suppose $M > 1/\la$, and suppose that $D$ is empty.  Then, for all $y \ge 0$, we have $\hat \phi_M(y) = u_M(y) = \min(My, 1)$.  By Proposition~\ref{prop:3.3}, $\hat \phi_M = u_M$ is a viscosity solution of the controller-stopper problem.  Because $M > 1/\la$, there exists $y_1 \in (1/M, \la)$.  The value function is identically 1 in a neighborhood of $y_1$, so that \eqref{eq:3.3} evaluated at $y=y_1$ becomes $\max [\la - y_1, 0] = 0$, which contradicts $y_1 < \la$. Thus, the region $D$ is non-empty.  
\end{proof}
In the next proposition, we provide a comparison result from which it follows that (together with Proposition~\ref{prop:3.1}) $\hat \phi_M$ is the {\it unique} viscosity solution of the controller-stopper problem. Our proof proof follows arguments that are used in the proof of Theorem 5.1 (pages 31-33) of Crandall et al.\ (1992) (see Touzi (2002) for a nice exposition on the viscosity solutions in the context of stochastic control problems). Since the controls are unbounded, the proof is a little more complicated. We use some of the techniques developed in the proofs of Theorem 4.2 in Duffie and Zariphopoulou (1993) and  Theorem 4.1 in Zariphopoulou (1994) to overcome these difficulties.
 In proving the comparison result, it will be more convenient to characterize the concept of viscosity solutions using parabolic semijets; see Crandall et al.\ (1992, Definition 2.2 and Remarks 2.3 and 2.4 (pages 10-11)).

{\lem \label{lem:3.5} Define the {\rm parabolic superjet} of  $v \in \mathcal{C}(\R_+)$ at $y \in \R_+$ by
$$ J^{2,+}v(y):=\{(p,X) \in \R^2: v(x) \le v(y)+ p(x - y)+ {1 \over 2} X (x - y)^2 + o\left(|x - y|^2\right) \; {\rm as}\; x \to y\}. $$
Also, define the {\rm parabolic subjet} of $v$ by $J^{2,-}v(y):=- J^{2,+}(-v(y))$. Then, $v \in \mathcal{C}(\R_+)$ is a viscosity supersolution $($subsolution$)$ of the controller-stopper problem if and only if for all $(p,X) \in J^{2,-} v(y)$ $($resp.\ $J^{2,+} v(y)),$ we have
\beq \label{eq:3.15}
\max  \bigg[ \la v(y) - y - (\la - \tilde r) y p - m y^2 X - \max_{\a} \left[ b \sqrt{1 - \rho^2} \, \a p + {1 \over 2} \a^2 X \right], 
 \; v(y) - u_M(y) \bigg] \ge ({\rm resp.}\leq) \; 0,
 \endeq}
 for all $y\in \R_+$.
{\prop \label{prop:3.6} 
Let $u$ and $v$ be nonnegative functions that are a viscosity subsolution and supersolution of \eqref{eq:3.15} on $\R_+$, respectively. If $u$ is continuous and $v$ is uniformly continuous on $\R_+$, then $u \leq v$ on $\rp$.
}

\begin{proof}  
We will prove the statement by a contradiction argument. 
Suppose that
\beq \label{eq:3.16}
C:=\sup_{x\in \R_+}(u(x)-v(x))> 0.
\endeq
\noi Observe that $C \leq 1$ because $u \leq u_M$, thanks to its subsolution property, and $v \geq 0$.

Given $\theta \in (0,1)$, we can choose an $\eps>0$ sufficiently small so that 
\[
\sup_{x \in \R_+}[u(x)-v(x)- \eps x^{\theta}]>0.
\]
Otherwise, there would exist a sequence $(\eps_n)_{n \in \mathbb{N}_+}$, with $\lim_{n \to \infty}\eps_n=0$, satisfying
\[
\sup_{x \in \R_+}[u(x)-v(x)- \eps_n x^{\theta}] \leq 0.
\]
But then
\[
0 \geq \sup_n \sup_{x \in \R_+}[u(x)-v(x)- \eps_n x^{\theta}] =  \sup_{x \in \R_+}\sup_n[u(x)-v(x)- \eps_n x^{\theta}] =  \sup_{x \in \R_+}[u(x)-v(x)],
\]
which contradicts \eqref{eq:3.16}.

Since $0 \leq u \leq 1$, $v \geq 0$, and $u$ and $v$ are both continuous, there exists a point
 $\tilde{x} \in \R_+$ such that
\beq \label{eq:maximizerofaux}
\sup_{x \in \R_+}[u(x)-v(x)- \eps x^{\theta}]=u(\tilde{x})-v(\tilde{x})- \eps \tilde{x}^{\theta}.
\endeq
For $\beta>0$, define
$$ \Psi(x,y):=u(x)-v(y)-{\beta \over 2}|x-y|^2-\eps x^{\theta}, \quad (x, y) \in \rp^2.$$
\noi The maximum of $\Psi$ is attained at a point $(\hat{x}(\beta, \eps),\hat{y}(\beta, \eps)) \in \rp^2$ 
at which we have
\beq \label{eq:3.17}
 {\beta \over 2}|\hat{x}(\beta, \eps)-\hat{y}(\beta, \eps)|^2+ \eps\hat{x}(\beta, \eps)^{\theta}< u(\hat{x}(\beta, \eps))-v(\hat{y}(\beta, \eps)) \leq 1,
\endeq
in which the first inequality follows from 
\beq \label{eq:psiatdifferent}
\Psi(\hat{x}(\beta, \eps),\hat{y}(\beta, \eps)) \geq \Psi(\tilde{x},\tilde{x})>0.
\endeq
Let us show that $\hat{x}(\beta, \eps) \neq 0$. First, note that $u(0)=0$, which follows from $0 \leq u \leq u_M$. If $\hat{x}(\beta, \eps)=0$, then \eqref{eq:3.17} would yield
\[
\frac{\beta}{2}\hat{y}(\beta, \eps)^2<-v(\hat{y}(\beta, \eps)).
\]
This gives us a contradiction because $v$ is a non-negative function.

For later use, we will now show that 
\beq \label{eq:betahatx}
\lim_{\beta \to \infty} \beta |\hat{x}(\beta, \eps)-\hat{y}(\beta, \eps)|^2 =0,
\endeq
and that 
\beq \label{eq:epsgoestozero}
\lim_{\eps \to 0} \lim_{\beta \to \infty} \eps \hat{x}(\beta, \eps)^{\theta}=0.
\endeq
The limit in \eqref{eq:betahatx} is a result of the following sequence of inequalities:
\beq \label{eq:modofcont}
\begin{split}
 {\beta \over 2}|\hat{x}(\beta, \eps)-\hat{y}(\beta, \eps)|^2 & = u(\hat{x}(\beta, \eps))-v(\hat{y}(\beta, \eps))- \eps \hat{x}(\beta, \eps)^{\theta} -\Psi(\hat{x}(\beta, \eps),\hat{y}(\beta, \eps))  \\
 & \leq u(\hat{x}(\beta, \eps))-v(\hat{x}(\beta, \eps))-\eps \hat{x}(\beta, \eps)^{\theta} -u(\tilde{x})+v(\tilde{x})+ \eps \tilde{x}^{\theta}+(v(\hat{x}(\beta, \eps))-v(\hat{y}(\beta, \eps))) \\& \leq w_{v} \left(\sqrt{{2  \over \beta}}\right),
\end{split}
\endeq
where $w_{v}$ is a modulus of continuity of the uniformly continuous function $v$. The first inequality in \eqref{eq:modofcont} follows from \eqref{eq:psiatdifferent}; and the second inequality is thanks to \eqref{eq:maximizerofaux} and \eqref{eq:3.17}.

Due to \eqref{eq:3.17}, $\eps \hat{x}(\beta, \eps)^{\theta}$ is bounded above uniformly in $\beta$. Hence (along a subsequence), we have that $\lim_{\beta \to \infty} \hat{x}(\beta, \eps)=x_0(\eps)$ for some $x_{0}(\eps) \in \R_+$. Letting $\beta \to \infty$ in \eqref{eq:psiatdifferent} yields
\begin{equation}
u(x_0(\eps))-v(x_0(\eps))- \eps x_0(\eps)^{\theta} \geq u(x)-v(x)- \eps x^{\theta}, \quad \text{for all $x \in \R_+$}.
\end{equation}
Taking the limit as $\eps \to 0$ (along a subsequence), we obtain that 
\[
\lim_{\eps \to 0}(u(x_0(\eps))-v(x_0(\eps)))- \lim_{\eps \to 0} \eps x_0(\eps)^{\theta} \geq u(x)-v(x), \quad \text{for all $x \in \R_+$},
\]
which implies that \eqref{eq:epsgoestozero} holds.

Again for future use, we will now analyze the parabolic superjet of $u$ at $\hat{x}(\beta, \eps)$ and the parabolic subjet of $v$ at $\hat{y}(\beta, \eps)$. We first apply Theorem 3.2 of Crandall et al.\ (1992) choosing $k=2$, $u_1(x):=u(x)-\eps x^{\theta}$, $u_2(x)=-v(y)$, and $\varphi(x,y)=(\beta/2) (x-y)^2$. In this case,
$\partial_x \varphi(\hat{x}(\beta, \eps),\hat{y}(\beta, \eps))=-\partial_y\varphi(\hat{x}(\beta, \eps),\hat{y}(\beta, \eps))=\beta(\hat{x}(\beta, \eps)-\hat{y}(\beta, \eps))$. As a result, $A=D^2 \varphi(\hat{x}(\beta, \eps),\hat{y}(\beta, \eps))$ is given by
\[
A=\beta \begin{pmatrix} 1 & -1 \\ -1 &1 \end{pmatrix},
\] 
which satisfies $A^2=2 \beta A$. Therefore, for every $\delta>0$, there exists
 a pair $(X,Y) \in \R^2$ such that 
 \[
 (\beta (\hat{x}(\beta, \eps)-\hat{y}(\beta, \eps)), X) \in {\bar{J}^{2,+}}u_1(\hat{x}(\beta, \eps)), \quad (\beta (\hat{x}(\beta, \eps)-\hat{y}(\beta, \eps)), Y) \in {\bar{J}^{2,-}}v(\hat{y}(\beta, \eps)),
 \]
 and 
\[
 - \left(\frac{1}{\delta}+2 \beta \right)\begin{pmatrix} 1 & 0 \\ 0 & 1 \end{pmatrix} \leq \begin{pmatrix} X & 0 \\ 0 & -Y \end{pmatrix} \leq (A+\delta A^2) = \beta(1+2 \delta \beta)  \begin{pmatrix} 1 & -1 \\ -1 & 1 \end{pmatrix}.
\] 
 (For two matrices $M$ and $N$, we write $M \geq N$ to mean that $M-N$ is positive semi-definite.) Here, $\bar{J}^{2,-}$ and $\bar{J}^{2,+}$ are defined as in page 11 of Crandall et al.\ (1992). 
 
 Choosing $\delta=1/\beta$, we obtain
\beq \label{eq:matineq}
 - 3 \beta \begin{pmatrix} 1 & 0 \\ 0 & 1 \end{pmatrix} \leq \begin{pmatrix} X & 0 \\ 0 & -Y \end{pmatrix} \leq 3 \beta  \begin{pmatrix} 1 & -1 \\ -1 & 1 \end{pmatrix}.
\endeq
It follows from \eqref{eq:matineq} that
\begin{equation}\label{eq:XY}
X \leq Y, \quad X, Y \in [-3 \beta, 3 \beta],
\end{equation}
Since for any $(b,B) \in {\bar{J}^{2,+}}u_1(\hat{x}(\beta, \eps))$, $(b+\eps \theta \hat{x}(\beta, \eps)^{\theta-1}, B+\eps \theta (\theta-1) \hat{x}(\beta, \eps)^{\theta-2}) \in  {\bar{J}^{2,+}}u(\hat{x}(\beta, \eps))$, we also have that
\begin{equation}
\begin{split}
& (\beta (\hat{x}(\beta, \eps)-\hat{y}(\beta, \eps))+ \eps \theta \hat{x}(\beta, \eps)^{\theta-1}, X+\eps \theta (\theta-1) \hat{x}(\beta, \eps)^{\theta-2}) \in {\bar{J}^{2,+}}u(\hat{x}(\beta, \eps)), \\&(\beta (\hat{x}(\beta, \eps)-\hat{y}(\beta, \eps)), Y) \in {\bar{J}^{2,-}}v(\hat{y}(\beta, \eps)).
 \end{split}
 \end{equation}

At this point we have gathered enough ammunition to contradict the assumption in \eqref{eq:3.16}. 
Let us denote 
\beq \label{eq:3.18}
F(x,p, Z)= x +(\la-\tilde{r}) x p+m x^2 Z +\max_{\a}\left[b \sqrt{1-\rho^2}\a p+ {1 \over 2} \a^2 Z\right]. 
\endeq
Either of the two cases holds for a given pair $(\beta,\eps)$ depending on the value of $Y$:
\begin{itemize}
\item[Case I.] When $Y>0$ or when both $Y=0$ and $\hat{x}(\beta, \eps)>\hat{y}(\beta, \eps)$, then
\[
F(\hat{y}(\beta, \eps), \beta (\hat{x}(\beta, \eps)-\hat{y}(\beta, \eps)), Y)=\infty.
\]
However, thanks to \eqref{eq:3.15} we have that
$v(\hat{y}(\beta, \eps)) \geq u_M(\hat{y}(\beta, \eps))$. On the other hand, since $u$ is a viscosity subsolution of \eqref{eq:3.15}, it necessarily satisfies $u(\hat{x}(\beta, \eps)) \leq u_{M}(\hat{x}(\beta, \eps))$. As a result we have
\beq \label{eq:case1}
0<u(\hat{x}(\beta, \eps))-v(\hat{y}(\beta, \eps)) \leq u_{M}(\hat{x}(\beta, \eps))-u_{M}(\hat{y}(\beta, \eps)).
\endeq
\item[Case II.] Otherwise, $F(\hat{y}(\beta, \eps), \beta (\hat{x}(\beta, \eps)-\hat{y}(\beta, \eps)), Y)<\infty$. Since $X \leq Y \leq 0$ and $\theta \in (0,1)$, we also have that
\[
 F(\hat{x}(\beta, \eps), \beta (\hat{x}(\beta, \eps)-\hat{y}(\beta, \eps))+\eps \theta \hat{x}(\beta, \eps)^{\theta-1}, X+\eps \theta (\theta-1) \hat{x}(\beta, \eps)^{\theta-2})<\infty.
\]
Using the supersolution property of $v$, the subsolution property of $u$, and the fact that $\max\{a,b\}-\max\{c,d\} \geq 0$ implies either $a \geq c$ or $b \geq d$, we obtain that either
\beq \label{eq:case2.1}
0 < u(\hat{x}(\beta, \eps))-v(\hat{y}(\beta, \eps)) \le u_M(\hat{x}(\beta, \eps)) - u_M(\hat{y}(\beta, \eps))
\endeq
\noi or
\beq \label{eq:case2.2}
\begin{split}
0&< 
 \la (u(\hat{x}(\beta, \eps))-v(\hat{y}(\beta, \eps)))
 \\&\leq F(\hat{x}(\beta, \eps),  \beta (\hat{x}(\beta, \eps)-\hat{y}(\beta, \eps))+\eps \theta \hat{x}(\beta, \eps)^{\theta-1}, X+\eps \theta (\theta-1) \hat{x}(\beta, \eps)^{\theta-2})-F(\hat{y}(\beta, \eps), \beta (\hat{x}(\beta, \eps)-\hat{y}(\beta, \eps)), Y)
 \\&= \hat{x}(\beta, \eps)-\hat{y}(\beta, \eps)+(\lambda-\tilde{r}) \beta (\hat{y}(\beta, \eps)-\hat{x}(\beta, \eps))^2+(\lambda-\tilde{r}) \theta \eps \hat{x}(\beta, \eps)^{\theta}+C_1+C_2,
 \end{split}
  \endeq
\noi in which
\beq \label{eq:A}
C_1= m \hat{x}(\beta, \eps)^2 X-m \hat{y}(\beta, \eps)^2 Y \leq 3 \beta m (\hat{x}(\beta, \eps)-\hat{y}(\beta, \eps))^2,
\endeq
 and
 \beq \label{eq:B}
 \begin{split}
 C_2&=\max_{\alpha}\left[\alpha b \sqrt{1-\rho^2}\left(\beta(\hat{x}(\beta, \eps)-\hat{y}(\beta, \eps))+\eps \theta \hat{x}(\beta, \eps)^{\theta-1}\right)+ {1 \over 2} \a^2 \left(X+\eps \theta (\theta-1)\hat{x}(\beta, \eps)^{\theta-2}\right)\right]\\&\quad -\max_{\alpha}\left[\alpha b \sqrt{1-\rho^2}\beta(\hat{x}(\beta, \eps)-\hat{y}(\beta, \eps))+ {1 \over 2} \a^2 Y\right]
 \\& \leq \max_{\alpha} \left[\alpha b \sqrt{1-\rho^2}\eps \theta \hat{x}(\beta, \eps)^{\theta-1}+ {1 \over 2} \a^2 \left(X-Y+\eps \theta (\theta-1)\hat{x}(\beta, \eps)^{\theta-2}\right)\right]
 \\&=-{1 \over 2} {\left(b \sqrt{1-\rho^2}\eps \theta \hat{x}(\beta, \eps)^{\theta-1}\right)^2 \over X-Y+\eps \theta (\theta-1)\hat{x}(\beta, \eps)^{\theta-2}} \leq -{b^2 (1-\rho^2) \theta \over 2 (\theta-1)} \eps \hat{x}(\beta, \eps)^{\theta}.
\end{split}
 \endeq
 The estimate in \eqref{eq:A} can be obtained by calculating
 \[
 \begin{pmatrix}\hat{x}(\beta, \eps) & \hat{y}(\beta, \eps) \end{pmatrix} \begin{pmatrix} X & 0 \\ 0 & -Y \end{pmatrix} \begin{pmatrix}\hat{x}(\beta, \eps) & \hat{y}(\beta, \eps) \end{pmatrix}'  \leq 3 \beta \begin{pmatrix}\hat{x}(\beta, \eps) & \hat{y}(\beta, \eps) \end{pmatrix}    \begin{pmatrix} 1 & -1 \\ -1 & 1 \end{pmatrix} \begin{pmatrix}\hat{x}(\beta, \eps) & \hat{y}(\beta, \eps) \end{pmatrix}',
 \]
\end{itemize}
Now, thanks to \eqref{eq:betahatx} and \eqref{eq:epsgoestozero}, the right-hand-sides of \eqref{eq:case1}, \eqref{eq:case2.1} and \eqref{eq:case2.2} all go to zero when we first let $\beta \to \infty$ and then $\eps \to 0$. This contradicts \eqref{eq:3.16}.   
\end{proof}

Since $\hat{\phi}_M$ is uniformly continuous and is bounded above by $u_M$, from Propositions \ref{prop:3.3} and \ref{prop:3.6} we deduce the following corollary:

{\cor \label{cor:3.7} $\hat \phi_M$ is the unique viscosity solution of \eqref{eq:3.3} among uniformly continuous positive functions.}

{\prop \label{prop:3.10} Assume that $M>1/\la$. Let $y_0<\infty$. The function $\hat \phi_M$ satisfies the smooth pasting condition, that is,
$$D_-\hat{\phi}_M(y_0)=0, \quad {\rm and} \quad D_+\hat{\phi}_M(y_M)=M, $$
in which $D_-$ denotes the left derivative operator and $D_+$ the right derivative operator.}

\pf The proof is motivated by the proof of Propositon 8.2 in Shreve and Soner (1994). We will prove the smooth pasting condition at $y_0$ (assuming that $y_0 < \infty$).  Smooth pasting at $y_M$ follows similarly.
Thanks to the concavity of $\hat{\phi}_M$, the right and the left derivatives exist at $y_0$.
Assume that 
\[
D_+\hat{\phi}_M(y_0)<D_{-}\hat{\phi}_M(y_0). 
\]
Let
\[
\delta \in (D_+\hat{\phi}_M(y_0), D_{-}\hat{\phi}_M(y_0))=(0,  D_{-}\hat{\phi}_M(y_0)). 
\]
Then, for any $\eps>0$
\[
\psi_{\eps}(y)=1+\delta(y-y_0)-{(y-y_0)^2 \over 2 \eps}, 
\]
dominates $\hat{\phi}_M$ locally at $y_0$. Since $\hat{\phi}_M$ is a viscosity subsolution of \eqref{eq:3.3} we have that
\beq \label{eq:3.32}
\lambda-y_0-(\la-\tilde{r})y_0 \delta+{m y_0^2 \over \eps } +{1 \over 2}b^2(1-\rho^2) {\delta^2 \over \eps} \leq 0. 
\endeq
Since \eqref{eq:3.32} can not hold for all $\eps>0$, our assumption that $D_+\hat{\phi}_M(y_0)<D_{-}\hat{\phi}_M(y_0)$ is not correct. We already know that $D_+\hat{\phi}_M(y_0)=0$; thus $D_{-}\hat{\phi}_M(y_0)=0$.
\end{proof}

{\prop \label{prop:3.8} For a point $\hat y \in D$, consider a neighborhood $N := (\hat y - \eps, \hat y + \eps) \subset D$ for some $\eps > 0$.  The function $y \to \hat{\phi}_M(y),$ $y \in [\hat{y}-\eps, \hat{y}+\eps],$ is the unique classical solution of the following non-linear boundary value equation:
\[
\begin{split}
& \la g = y + (\la - \tilde r) y g' + m y^2 g'' + \max_\a \left[ b \sqrt{1 - \rho^2} \, \a \, g' + {1 \over 2}
\a^2 g'' \right] \; \hbox{ on } N, \cr
& g(\hat y - \eps) = \hat \phi_M(\hat y - \eps) \; \hbox{ and } \; g(\hat y + \eps) = \hat \phi_M(\hat y + \eps),
\end{split}
\]
Moreover, $y \to \hat{\phi}_M(y)$ is strictly increasing and concave on $N$.}

\pf  (i) First, assuming that $y \to \hat{\phi}_M(y)$, $y \in D$, is $\mathcal{C}^2$, we will show that $\hat{\phi}_M''(y)<0$, $y \in D$. Since $y \to \hat{\phi}_M(y)$ is concave, we already have that  $\hat{\phi}_M''(y) \leq 0$, $y \in D$. Because $y \to \hat{\phi}_M(y)$ satisfies \eqref{eq:3.15} with inequality $\geq$, it is clear that $\hat{\phi}_M''(y)<0$ for any $y \in D$ satisfying $\hat{\phi}'_M(y)>0$. We only need to show that there is no $y \in D$ such that  
$\hat{\phi}_M''(y)=0$ and $\hat{\phi}'_M(y)=0$. Let $\tilde{y}\in \R_+$ be the smallest such point;
then, it would satisfy $\hat{\phi}_M(\tilde{y})=\tilde{y}/\lambda$, again due to \eqref{eq:3.15}. That is, $\tilde{y}$ lies at the intersection of $y \to y/\lambda$ and $y \to \hat{\phi}_M(y)$. Moreover, $\tilde{y} \leq \lambda$ since $\hat{\phi}_M(y) \leq u_M(y)$ for all $y \in \R_+$. If $\tilde{y}=\lambda$, then our claim that  $\hat{\phi}_M''(y)<0$ for all $y \in D$ holds since in that case $u_M(y)=\hat{\phi}_M(y)$, $y \geq \tilde{y}=\lambda$; i.e., $\lambda$ is an element of the stopping region and not an element of $D$.

Thus, $\tilde{y}<\lambda$. Since $\hat{\phi}_M$ is concave and nondecreasing, $\hat{\phi}_M'(y)=0$ for $y>\tilde{y}$. Therefore, $\hat{\phi}_M(y)=\tilde{y}/\lambda$ for $y \geq \tilde{y}$. Let $\hat{y}\in D$ be such that $\hat{y}>\tilde{y}$. We have that $\hat{\phi}_M''(\hat{y})=0$ and $\hat{\phi}'_M(\hat{y})=0$. According to \eqref{eq:3.15}, $\hat{y}$ should satisfy $\hat{\phi}_M(\hat{y})=\hat{y}/\lambda$, which contradicts our observation that $\hat{\phi}_M(\hat{y})=\tilde{y}/\lambda$, and hence contadicts $\tilde{y}<\lambda$.

(ii) Since $\hat{\phi}_M''(y)<0$, $y \in D$, and $\hat{\phi}_M'(y_0)=0$, we must have that $\hat{\phi}_M'(y)>0$ for $y \in D$.

(iii) In the rest of the proof, we will show that $\hat{\phi}_M$ is a classical solution of the above non-linear boundary value problem. The proof follows steps that are similar to the ones in the proof of Theorem 5 in Duffie et al.\ (1997). We will point out only the necessary modifications.   For $L \in \zp$, define

\[
w^L(y) = \inf_\tau \sup_{\a \in {\mathcal A}(y), -L \leq \a \leq L} \hat \E^y \left[ \int_0^{\tau} e^{-\la t} \, Y^\a_t \, dt
+ e^{-\la \tau} \, u_M(Y^\a_\tau) \right]. 
\]
\noi Observe that $w^L$ is concave, which follows from the same line of argument as for the concavity of $\hat \phi_M$. One can show (as in Propositions~\ref{prop:3.3} and \ref{prop:3.6},  and Corollary \ref{cor:3.7}) that $w^L$ is the unique viscosity solution of
\[
 \left[ \la g - y - (\la - \tilde r) y ' - m y^2 g'' - \max_{-L \leq \a \leq L} \left[ b \sqrt{1 - \rho^2} \, \a g' + {1 \over 2} \a^2 g'' \right], g - u_M \right]  = 0
 \]
\noi among positive concave functions bounded above by $u_M$. Note that $w^L$ satisfies $|w^{L}(x)-w^{L}(y)| \leq M |x-y|$ for all $x, y \in \rp$; that is, $\{w^L\}$ is an equicontinuous sequence of functions. Since each $w^L$ is bounded above by 1, the increasing sequence $\{w^{L}\}$ converges to some function, say $\hat{w}$, which is continuous (continuity follows from the equicontinuity of the approximating sequence).  Now from Dini's theorem (since $\{w^L\}$ is an increasing sequence of continuous functions converging to a continuous function), we have that $\{w^L\}$ converges to $\hat{w}$ uniformly on compact sets.  Then for any $x,y \in \R_+$ and for any $\eps > 0$, there exists an $L(\eps)$ such that for any $L \geq L(\eps)$, $\|w^L\|_{\infty} > \|\hat{w}\|_{\infty} - \eps$. We already know that $\hat w > w^L$ on $\rp$. Together with the concavity of $w^L$, this leads to
\[
\begin{split}
\hat{w}(\gamma x+(1- \gamma y)) &\geq \gamma w^L(x)+(1 - \gamma)w^{L}(y) \\
& \ge \gamma \hat w(x) + (1 - \gamma) \hat w(y) - \eps,
\end{split}
\]
\noi for $\gamma \in [0, 1]$ and $x, y \in \rp$.  Because $\eps$ is arbitrary, this implies that $\hat w$ is concave. Therefore $w^L$ converges locally uniformly to a concave function $\hat{w}$.

From Theorem I.3 of Lions (1983) (a result on the stability of viscosity solutions), it follows that $\hat{w}$ is a viscosity
solution of (2.13).  Note that $\hat{w}$ is positive concave, and is bounded by $u_M$. On the other hand, from Proposition 3.6, it follows that (2.13) has a unique viscosity solution among uniformly continuous, concave functions. Therefore, from
Corollary 3.7, we conclude that $w^L \to \hat{\phi}_M$ locally uniformly in $(0, \infty)$.

Since $D$ defined in \eqref{eq:3.2} is a subset of the corresponding set for $w^L$, namely $D^L := \{ y \in \rp: w^L(y) < u_M(y) \}$, it follows that $w^L$ is a viscosity solution of
\beq \label{eq:3.21}
\begin{split}
& \la g = y + (\la - \tilde r) y g' + m y^2 g'' + \max_{-L \leq \a \leq L} \left[ b \sqrt{1 - \rho^2} \, \a \, g' +
{1 \over 2} \a^2 g'' \right] \; \hbox{ on } N, \cr
& g(\hat y - \eps) = w^L(\hat y - \eps) \; \hbox{ and } \; g(\hat y + \eps) =w^L(\hat y + \eps).
\end{split}
\endeq
\noi By using Theorem II.1 of Lions (1983), one can prove that, indeed, $w^L$ is the unique viscosity solution of \eqref{eq:3.21}. The rest of our proof follows the same arguments after equation (6.3) in the proof of Theorem 5 in Duffie et al.\ (1997). \end{proof}

\begin{cor}\label{cor:3.11}
The value function $\hat{\phi}_M$ is a classical solution of \eqref{eq:2.15}. 
\end{cor}

\begin{proof}
Since  $\hat{\phi}_M(y) =u_M(y)$ for $y \notin D$, the claim is a corollary of Proposition~\ref{prop:3.8}.


\end{proof}

The following result shows that \eqref{eq:2.13} has a saddle point.

{\prop \label{prop:3.9} 

(i) Let us define  $\a^*:D \to \R_+$ by
\beq \label{eq:3.23}
 \alpha^*(y) = -b \sqrt{1-\rho^2} \; {\hat{\phi}_M'(y) \over \hat{\phi}_M''(y)}.
 \endeq
This function is locally Lipschitz and satisfies $0 \leq \alpha(y) \leq C y$, $y \in D$, for some positive constant $C$. 

(ii) Extend $\alpha^*$ from $D$ to $\R_+$ in such a way that it still is locally Lipschitz, it has linear growth, and $\alpha^*(0)=0$. Let $Y^*$ denote the diffusion 
whose dynamics are given by
\begin{equation}\label{eq:Ystar}
dY^*_t=Y_t^*\left[(\lambda-\tilde{r})dt+{\mu-r-\sigma b \rho \over \sigma}d \hat{B}_t^{(1)}\right]+ \alpha^*(Y_t^*) \left[b
\sqrt{1-\rho^2}dt+ d \hat{B}_t^{(2)}\right],\quad Y^*_0=y.
\end{equation}
This stochastic differential equation has a unique strong solution. Moreover, $\hat \E[\int_0^{t}(\a^*(Y^*_s))^2ds]<\infty$ for all $t \in \R_+$.

(iii) Define $\tau^*(Y^{y,\a})=\inf\{t \geq 0: Y_t^{y,\a} \notin D\}$. Then, $\hat{\phi}_M$ is the unique classical solution of \eqref{eq:2.15} and satisfies
\beq \label{eq:3.24}
\hat{\phi}_M(y)={\hat \E} \left[\int_0^{\tau^*(Y^*)}e^{-\la t} Y_t^* dt+ e^{-\la \tau^*(Y^*)}u_M \left(Y^*_{\tau^*(Y^*)} \right)\right]. 
\endeq
Moreover,
\beq \label{eq:3.25}
\begin{split}
{\hat \E} \left[\int_0^{\tau^*(Y^{y,\a})}e^{-\la t} Y_t^{y,\a} dt+ e^{-\la \tau^*(Y^{y,\a})}u_M \left(Y^{y,\a}_{\tau^*(Y^{y,\a})} \right)\right] & \leq \hat{\phi}_M(y)  \leq {\hat \E} \left[\int_0^{\tau}e^{-\la t} Y_t^* dt+ e^{-\la \tau}u_M(Y^*_{\tau})\right],
\end{split}
\endeq
for any stopping time $\tau$ of the filtration generated by $B^{(1)}$ and $B^{(2)}$ and for any admissible strategy $\a$, which implies that the the controller stopper problem satisfies the min-max principle, i.e.,
\[
\hat \phi_M(y) = \sup_{\a \in \mathcal{A}(y)} \inf_\tau  \hat \E \left[ \int_0^{\tau} e^{-\la t} \, Y^{y,\a}_t \, dt + e^{-\la \tau} \,
u_M(Y^{y,\a}_\tau) \right]=\inf_\tau  \sup_{\a \in \mathcal{A}(y)}  \hat \E \left[ \int_0^{\tau} e^{-\la t} \, Y^{y,\a}_t \, dt + e^{-\la \tau} \,
u_M(Y^{y,\a}_\tau) \right],
\]
and that the pair $(\tau^*,\alpha^*)$ is a saddle point.
}

\pf 
(i) Using the fact that $\hat{\phi}_M$ satisfies \eqref{eq:2.15} and that it is strictly concave and strictly increasing on $D$ we can write the function $\alpha^*$ on $D$ as 
\begin{equation}\label{eq:alpha-star}
\begin{split}
\alpha^*(y)&= \frac{1}{b \sqrt{1-\rho^2}} \left\{-\left[\frac{y-\la \hat{\phi}_M(y)}{\hat{\phi}_M'(y)}+(\la-\tilde{r})y\right]+\sqrt{\left[\frac{y-\la \hat{\phi}_M(y)}{\hat{\phi}_M'(y)}+(\la-\tilde{r})y\right]^2+2 b^2 (1-\rho^2)m y^2}\right\} 
\\&\leq \sqrt{2m} y.
\end{split}
\end{equation}
The first line shows that $y \to \alpha^*(y)$, $y \in D$, is locally Lipschitz. The second line shows that the same function has linear growth.

(ii) The SDE for $Y^*$ has a unique strong solution since $\alpha^*$ on $\R_+$ is locally Lipschitz and has linear growth. (Note that \eqref{eq:alpha-star} can be used to extend $\alpha$ from $D$ to $\R_+$ for values of $y$ that are close to zero.) 

The fact that $\hat \E[\int_0^{t}(\a^*(Y^*_s))^2ds]<\infty$ for all $t \in \R_+$, holds since $\a^*$ satisfies the linear growth condition in (i). The proof follows from Lemma 11.5 on page 129 of Rogers and Williams (2000) and Gronwall's inequality.

(iii)
Next, \eqref{eq:3.24} follows from applying It\^{o}'s formula to $e^{-\la t} \, \hat{\phi}_M(Y_t^*)$ and by using the fact that $\hat{\phi}_M$ satisfies \eqref{eq:2.15}. In fact, thanks to It\^{o}'s formula, any solution of \eqref{eq:2.15} can be represented by the right-hand-side of \eqref{eq:3.24}.

To show the second inequality in \eqref{eq:3.25}, we will argue that the function $\eta$ defined by
\beq \label{eq:3.26}
 \eta(y) := \inf_{\tau} {\hat \E} \left[\int_0^{\tau}e^{-\la t} Y_t^* dt+ e^{-\la \tau} u_M(Y^*_{\tau})\right], \endeq
\noi is equal to $\hat{\phi}_M$ on $\rp$; specifically, the infimum in \eqref{eq:3.26} is attained at $\tau^*(Y^*)$.  To this end, define a process $X$ by
\beq \label{eq:3.27}
 X_t := \int_0^{t}e^{-\la s} Y_s^* ds+ e^{-\la t}u_M(Y^*_{t}). 
 \endeq
\noi By using the strong Markov property of $Y^*$, we can write the Snell envelope $\xi$ of $X$ as
\beq \label{eq:3.28}
 \xi_t := \inf_{\tau \geq t} \hat{\E} \{ X_{\tau} \big| \hat{\mathcal{F}}_t \} = \int_0^{t}e^{-\la s}Y^*_s ds + e^{-\la t}\eta(Y_t^*).
 \endeq
\noi The derivation of this equation is similar to the derivation of equation (7.6) in Chapter 2 of Karatzas and Shreve (1998). Now, from Theorem D.12 in Karatzas and Shreve (1998), it follows that the stopping time
$$ \tilde{\tau} := \inf\{t \geq 0: \xi_t=X_t\} $$
\noi is optimal.  Note, by \eqref{eq:3.27} and \eqref{eq:3.28}, that $\tilde{\tau}=\tau^*(Y^*)$, which proves that $\eta = \hat{\phi}_M$.

We now prove the first inequality in \eqref{eq:3.25}. By applying It\^{o}'s formula to $e^{-\la t} \, \hat{\phi}_M(Y_t^{\a})$, we get
\[
\begin{split}
&\hat{\phi}_M(y)={\hat \E} \left[e^{-\la \tau^*(Y^{y,\a})}\hat{\phi}_M \left(Y^{y,\a}_{\tau^*(Y^{y,\a})} \right)\right]+ \cr
&{\hat \E}\left[\int_0^{\tau^*(Y^{y,\a})} e^{-\la t} \left( \la \hat{\phi}_M(Y_t^\a) - \hat{\phi}_M' (Y_t^\a)(Y_t^\a (\la-\tilde{r}) + \a_t b \sqrt{1-\rho^2})- \hat{\phi}_M''(Y_t^\a) \left(m + {1 \over 2} \a_t^2 \right) \right) dt \right] \cr
& \geq  {\hat \E} \left[\int_0^{\tau^*(Y^{y,\a})}e^{-\lambda t} Y^{y,\a}_t dt+ e^{-\lambda \tau^*(Y^{y,\a})} u_M \left(Y^{y,\a}_{\tau^*(Y^{y,\a})} \right)\right],
\end{split}
\]
\noi in which the inequality follows from the definition of $\tau^*(Y^{y,\a})$ and the fact that  $\hat{\phi}_M$ satisfies \eqref{eq:2.15}.   \end{proof}
\subsection{Proofs of Theorems~\ref{thm:2.2} and \ref{thm:2.3}}\label{sec:3.3}

In this section, we prove Theorems \ref{thm:2.2} and \ref{thm:2.3} through a series of propositions as outlined in items 6 through 10 in Section~\ref{sec:3.1}.  First, we define a second-order differential operator associated with the minimization problem in \eqref{eq:2.7} as follows:  For an open set $G \subset (0, M)$, $v \in \mathcal{C}^2(G)$, and $\a \in \R$, define the function $\mathcal{L}^\a v: G \to \bf R$ by
\beq \label{eq:3.33}
\begin{split}
\mathcal{L}^{\a} v(z) &= - \la v(z) + \left[  \left( \tilde r z - 1 \right) + \left( \mu - r - \sigma b \rho \right) \a \right] v'(z) + {1 \over 2} \left[ b^2 (1 - \rho^2) z^2 + \sigma^2 \a^2 \right] v''(z).
\end{split}
\endeq
\noi We have the following verification lemma that shows that a suitably smooth solution of \eqref{eq:2.8} equals $\phi_M$, with the optimal investment strategy given in \eqref{eq:2.9}.
{\lem \label{lem:3.12}Suppose the real-valued functions $v$ on $\rp$ and $\beta$ on $(0, M)$ satisfy the following
conditions:
\begin{itemize}
\item[(0)] $v$ is continuous and non-increasing on $\rp;$
\item[(i)] $v \in \mathcal{C}^2(\rp - \{M \});$
\item[(ii)] $\min_{\a}\mathcal{L}^{\a} v(z) = \mathcal{L}^{\beta(z)} v(z)=0$;
\item[(iii)] $v(0) = 1$ and $v(z) = 0$ for $z \ge M$.
\end{itemize}
{\it Under the above conditions, the modified minimum probability of the lifetime ruin $\phi_M$ in $\eqref{eq:2.7}$ is given by} 
\beq \label{eq:3.34}
\phi_M(z) = v(z), \quad z \in \rp.
\endeq}

\pf  For an arbitrary strategy $\tilde \pi \in \widetilde{\mathcal{A}}$, let $Z^{\tilde \pi}$ denote the wealth process when we use $\tilde \pi$ as the investment policy.  Recall the hitting times $\tilde \tau^z_0 = \inf \{ t \ge 0: Z^{\tilde \pi}_t \le 0 \}$ and $\tilde \tau^z_M = \inf \{ t \ge 0: Z^{\tilde \pi}_t \ge M \}$. (Technically, we should apply the superscript $\tilde \pi$ to the stopping times, but we omit it because the notation is otherwise too cumbersome.)  Because the time of death of the individual $\tau_d$ is independent of the Brownian motions $B^{(1)}$ and $B^{(2)}$, we can write $\phi_M$ as
\beq \label{eq:3.36}
\begin{split}
\phi_M(z) &= \inf_{\tilde \pi \in \widetilde{\mathcal{A}}} \tilde {\bf E} \int_0^\infty \la e^{-\la s} \, {\bf 1}_{\{ \tilde \tau_0^z < \tilde \tau^z_M \wedge s \}} \, ds \cr
&= \inf_{\tilde \pi \in \widetilde{\mathcal{A}}} \tilde {\bf E} \int_{\tilde \tau^z_0}^\infty \la e^{-\la s} \, {\bf 1}_{\{ \tilde \tau^z_0 < \tilde \tau^z_M \}} \, ds = \inf_{\tilde \pi \in \widetilde{\mathcal{A}}} \tilde {\bf E} \left( e^{-\la \tilde \tau^z_0} \, {\bf 1}_{\{ \tilde \tau^z_0 < \tilde \tau^z_M \}} \right).
\end{split}
\endeq
\noi By using this formulation of the problem, the verification lemma follows from classical arguments, as we proceed to demonstrate.  First, for any positive integer $n$, define the stopping time $\tilde \tau_n$ by $\tilde \tau_n = \inf \{ t \ge 0: \int_0^t  \tilde \pi^2_s \, ds \ge n \} \wedge \inf \{ t \ge 0: \int_0^t  (Z^{\tilde \pi}_s - \tilde \pi_s)^2 \, ds \ge n \} \wedge \inf \{ t \ge 0: Z^{\tilde \pi}_t \le 1/n \}$.  Then, define the stopping time $\tilde \tau^{(n)} = \tilde \tau^z_0 \wedge \tilde \tau_n \wedge \tilde \tau^z_M$.

Assume that we have the function $v$ as specified in the statement of this lemma.  By applying It\^o's formula to the function $f$ given by $f(z, t) = e^{-\la t} \, v(z)$, we have
\beq \label{eq:3.37}
\begin{split}
& e^{-\la \left( t \wedge \tilde \tau^{(n)} \right)} v \left(Z^{\tilde \pi}_{t \wedge \tau^{(n)}} \right) = v(z) - \la \int_0^{t \wedge \tilde \tau^{(n)}} e^{-\la s} \, v(Z^{\tilde \pi}_s) \, ds \cr
& \quad + \int_0^{t \wedge \tilde \tau^{(n)}} e^{-\la s} \left( \left( \tilde r Z^{\tilde \pi}_s - 1 \right) + \left( \mu - r - \sigma b \rho \right) \tilde \pi \right) v'(Z^{\tilde \pi}_s) \, ds \cr
& \quad + {1 \over 2} \int_0^{t \wedge \tilde \tau^{(n)}} e^{-\la s} \left( b^2 (1 - \rho^2) (Z^{\tilde \pi_s})^2 + \sigma^2 \tilde \pi^2 \right) v''(Z^{\tilde \pi}_s) \, ds \cr
& \quad + \int_0^{t \wedge \tilde \tau^{(n)}} e^{-\la s} \, v'(Z^{\tilde \pi}_s) \, \left( b \sqrt{1 - \rho^2} \, (Z^{\tilde \pi}_s - \tilde \pi_s)  \, d\tilde B^{(1)}_s + \tilde \pi_s \sqrt{b^2 (1 - \rho^2) + \sigma^2} \, d\tilde B^{(2)}_s \right)   \cr
& = v(z) + \int_0^{t \wedge \tilde \tau^{(n)}} e^{-\la s} \, \mathcal{L}^{\tilde \pi_s} v(Z^{\tilde \pi}_s) ds \cr
& \quad + \int_0^{t \wedge \tilde \tau^{(n)}} e^{-\la s} \, v'(Z^{\tilde \pi}_s) \, \left( b \sqrt{1 - \rho^2} \, (Z^{\tilde \pi}_s - \tilde \pi_s)  \, d\tilde B^{(1)}_s + \tilde \pi_s \sqrt{b^2 (1 - \rho^2) + \sigma^2} \, d\tilde B^{(2)}_s \right), 
\end{split}
\endeq
\noi in which the second equality follows from the definition of $\mathcal{L}^{\a}$ in \eqref{eq:3.33}.

If we take the expectation of both sides, the expectation of the last term in \eqref{eq:3.37} is zero because
\[
\begin{split}
&\tilde {\bf E} \left[ \int_0^{t \wedge \tilde \tau^{(n)}} e^{-2 \la s} \left( b^2 (1 - \rho^2) (Z^{\tilde \pi}_s - \tilde \pi_s)^2 + (b^2 (1 - \rho^2) + \sigma^2) \, \tilde \pi_s^2 \right) (v'(Z^{\tilde \pi}_s))^2  \, ds \right] \cr
& \quad \le \max_{z \in [1/n, M]}  (v'(z))^2 \left( (b^2 (1 - \rho^2) + \sigma^2) \tilde \E \left[ \int_0^{t \wedge \tilde \tau^{(n)}} \left( (Z^{\tilde \pi}_s - \tilde \pi_s)^2 + \tilde \pi_s^2 \right) ds \right]   \right) < \infty,
\end{split}
\]
\noi because $v'(z)$ is bounded on $[1/n, M]$ and because of the definition of $\tilde \tau_n$.  Thus, we have
\beq \label{eq:3.38}
\tilde {\bf E} \left[e^{-\la \left( t \wedge \tilde \tau^{(n)} \right)} v(Z^{\tilde \pi}_{t \wedge \tau^{(n)}}) \right] = v(z) +\tilde {\bf E}\left[ \int_0^{t \wedge \tilde \tau^{(n)}} e^{-\la s} \, \mathcal{L}^{\tilde \pi_s} v(Z^{\tilde \pi}_s) ds \right] \ge v(z), 
\endeq
\noi where the inequality follows from assumption (ii) of the proposition.

Because $v$ is bounded, $v(0) = 1$, and $v(M) = 0$, it follows from \eqref{eq:3.38} and the dominated convergence theorem that
\beq \label{eq:3.39}
v(z) \le\tilde {\bf E}\left( e^{-\la \tilde \tau^z_0} v(Z^{\tilde \pi}_{\tilde \tau^z_0}) \, {\bf 1}_{\{ \tilde \tau^z_0 < \tilde \tau^z_M \}} \right) =\tilde {\bf E} \left(e^{-\la \tilde \tau^z_0} \, {\bf 1}_{\{ \tilde \tau^z_0 < \tilde \tau^z_M \}} \right), 
\endeq
\noi for any $\pi \in \widetilde{\mathcal{A}}$.  Thus, it follows from \eqref{eq:3.36} that $v \le \phi_M$.

Now, let $\beta$ be as specified in the statement of this lemma; that is, $\beta$ is the minimizer of $\mathcal{L}^{\tilde \pi} v$.   It
follows from the above argument that we will have equality in \eqref{eq:3.39}, from which it follows that $v = \phi_M$. \end{proof}

The following proposition follows easily from Lemma~\ref{lem:3.12}:

{\prop \label{prop:3.13}The Legendre transform of $\hat \phi_M$ solves the HJB equation \eqref{eq:2.8}  on $[0, M]$ and thereby equals the minimum probability of ruin $\phi_M$. As a result, $\phi_M$ is strictly decreasing on $[0,M]$ and strictly convex on $(0,M)$.
Also, the optimal investment strategy is given by the expression in $\eqref{eq:2.9}$.}
\pf As we showed in Theorem~\ref{thm:2.5}, the Legendre transform $\Phi_M$ of $\hat \phi_M$ given \eqref{eq:2.18} satisfies the conditions given in Lemma~\ref{lem:3.12}.  This proves that $\Phi_M=\phi_M$. Recall the convexity and the monotonicity properties of $\Phi_M$ from Theorem~\ref{thm:2.5}.

Let us define $\beta(z) := - {\mu - r - \sigma b \rho \over \sigma^2} {\phi'_M(z) \over \phi''_M(z)}$ for $z \in (0, M)$. This function minimizes $\mathcal{L}^{\a} \hat{\phi}_M(z)$ over $\a$; hence it is a candidate optimal strategy.
To conclude the optimality of this strategy we need to show that $\beta$ is locally Lipschitz, which implies that \eqref{eq:Z-dyn} (with $\tilde{\pi}_t$ replaced by $\beta(Z_t)$) has a unique strong solution up to the first time $\tau$ such that $Z_{\tau}$ is equal to either $0$ or $M$.

Using the fact that $\phi_M$ solves \eqref{eq:2.5}, we can write
\[
\begin{split}
\beta(z)&=\frac{1}{\mu-r-\rho \sigma b}\left[-\left(\frac{-\la \phi_M}{\phi_M'}+\tilde{r}z-1\right)+\sqrt{\left(\frac{-\la \phi_M}{\phi_M'}+\tilde{r}z-1\right)^2+ \left(\frac{\mu-r-\rho b \sigma}{\sigma^2}\right)^2b^2 (1-\rho^2)z^2}\right], 
\end{split}
\]
which shows that $\beta$ is indeed locally Lipschitz, since it is a continuously differentiable function.
  \end{proof}

In the next sequence of propositions, we prove Theorem~\ref{thm:2.2} and that $\lim_{M \to \infty} \phi_M = \phi$ on $\rp$.
{\prop \label{prop:3.14} Define $\tilde \phi$ on $\rp$ by
\beq \label{eq:3.40}
 \tilde{\phi}(z)=\lim_{M \rightarrow \infty}\uparrow \phi_M(z). 
 \endeq
Then, $\tilde{\phi}(0)=1$, $\tilde{\phi}$ is convex and it is a viscosity solution of $\eqref{eq:2.5}$. Moreover, the convergence in \eqref{eq:3.40} is uniform.}
\pf Since $\phi_M(0)=1$ for all $M$, it follows that $\tilde{\phi}(0)=1$.  It immediately follows that $\tilde{\phi}$ is convex since it is the upper envelope of convex functions; that is, $\tilde \phi(z)=\sup_{M} \phi_M(z)$ for $z \in \rp$.

Since $\{ \phi_M(z) \}$ is increasing with respect to $M > 0$ for all $z \in \rp$, we can apply Dini's theorem and conclude that $\phi_M$ converges to $\tilde{\phi}$ uniformly on compact sets of $\R_+$. Below, we will show that $\tilde{\phi}$ is a viscosity subsolution of \eqref{eq:2.5}. The fact that it is a viscosity supersolution of \eqref{eq:2.5} can be similarly proved.

Define $F$ by
\beq \label{eq:3.41}
\begin{split}
F(z,u(z), u'(z), u''(z)) &= \lambda u(z) - (\tilde{r}z-1) u'(z) - {1 \over 2}b^2 (1-\rho^2) z^2 u''(z) \cr
& \quad -\min_{\tilde{\pi}}\left[(\mu-r-\sigma\, b\, \rho)\tilde{\pi}u'(z) + {1 \over 2}\sigma^2 \tilde{\pi}^2 u''(z) \right], 
\end{split}
\endeq
\noi for a test function $u \in \mathcal{C}^2 (\R_+)$ and for $z \in \rp$.  Note that $F$ is non-increasing with respect to its fourth argument $u''(z)$. For $O \subset \R_+$ open, let $\psi \in \mathcal{C}^2(O)$, and suppose $\tilde{\phi}-\psi$ has a strict local maximum at $z_0 \in O$. We will show that
\beq \label{eq:3.42}
 F(z_0, \tilde{\phi}(z_0), \psi'(z_0), \psi''(z_0)) \leq 0,
\endeq 
\noi and conclude by using Remark I.9 in Lions (1983). If $\delta>0$ is small enough, $[z_0-\delta, z_0+\delta] \subset O$ and
$$ (\tilde{\phi}-\psi)(z_0) >\max \{(\tilde{\phi}-\psi)(z_0-\delta), (\tilde{\phi}-\psi)(z_0+\delta)\}. $$
\noi Since $\{ \phi_M \}$ converges to $\tilde{\phi}$ uniformly on compact sets, we can choose $M=M(\delta) \; (>z_0+\delta)$ large enough so that
$$ \max_{z \in [z_0-\delta, z_0+\delta]}(\phi_M-\psi)(z)> \max
\{(\phi_M-\psi)(z_0-\delta), (\phi_M-\psi)(z_0+\delta)\}. $$
\noi As a result, there exists $z_{\delta} \in (z_0-\delta,z_0+\delta)$ such that
\beq \label{eq:3.43}
 \max_{z \in [z_0-\delta, z_0+\delta]}(\phi_M-\psi)(z)=(\phi_M-\psi)(z_\delta). 
\endeq
\noi Thus, $z_\delta$ is a local maximum of $\phi_M - \psi$, from which we conclude that $\phi'_M(z_\delta) = \psi'(z_\delta)$ and $\phi''_M(z_\delta) \leq \psi''(z_\delta)$.
From Proposition~\ref{prop:3.13}, we know that $\phi_M$ is a smooth solution of \eqref{eq:2.8} on $[0, M]$.  Thus, 
\[F(z, \phi_M(z),\phi_M'(z), \phi_M''(z)) \leq 0
\] for $z \in [z_0-\delta,z_0+\delta]$ (recall that $M(\delta)>z_0+\delta$).  It follows from \eqref{eq:3.43} that
$$ F(z_{\delta}, \phi_M(z_{\delta}),\psi'(z_{\delta}), \psi''(z_{\delta})) \leq 0,$$
\noi because $F$ is non-increasing with respect to its fourth argument.

Observe that as $\delta \to 0$, we have $z_{\delta} \rightarrow z_0$ and $\phi_M(z_{\delta}) \rightarrow \tilde{\phi}(z_0)$. Moreover, since $\psi \in \mathcal{C}^2(O)$, it follows that $\psi'(z_\delta) \rightarrow \psi'(z_0)$ and $\psi''(z_{\delta}) \rightarrow \psi''(z_0)$. Finally, the continuity of $F$ implies that \eqref{eq:3.42} holds.  \end{proof}

{\prop \label{prop:3.15} The function $\tilde{\phi}$ given in \eqref{eq:3.40} is a smooth solution of $\eqref{eq:2.5}$.}

\pf 
Due to the convexity of $\tilde{\phi}$ we can choose points $z_1<z_2$ such that the derivative of $\tilde{\phi}$ at points $z_1$ and $z_2$ exists. (Also, recall that $\tilde{\phi}$ is almost everywhere differentiable.)

For a given positive $h < (z_2-z_1)/2$, we can find a sufficiently large $M$ such that
\begin{equation}\label{eq:unbd}
0 \leq \tilde{\phi}(z)-\phi_M(z) \leq h, \quad z \in \R_+,
\end{equation}
thanks to Proposition~\ref{prop:3.14}.
Using the convexity of $\phi_M$ and $\tilde{\phi}$, we deduce that
\[
\phi_M'(z) \geq \frac{\phi_M(z)-\phi_M(z-h)}{h} \geq \frac{\tilde{\phi}(z)-\tilde{\phi}(z-h)}{h}-1 \geq C_1:=\tilde{\phi}'(z_1)-1,
\]
for any $z \geq z_1+h$. On the other hand,
\[
\tilde{\phi}'(z_2) \geq \frac{\tilde{\phi}(z_2)-\tilde{\phi}(z_2-h)}{h} \geq \frac{\phi_M(z_2)-\phi_M(z_2-h)}{h}+1 \geq \phi_M'(z_2-h)+1,
\]
which implies that
\[
C_2:=\tilde{\phi}'(z_2)-1 \geq \phi'_M(z_2-h) \geq \phi_M'(z),
\]
for $z \leq z_2-h$. Since $C_1$ and $C_2$ do not depend on $h$, which can be taken to be arbitrarily small, we have that
\[
C_1 \leq \phi_M'(z) \leq C_2, \quad z \in (z_1,z_2).
\]

Since $\phi_M$ is decreasing and convex, we have that $\phi_M'(0)  \leq \phi'_M(z) \leq 0$ for all $z \in [0,M]$. Next, for $z <z_2$, we will show that $\phi_M''(z) > K(z) > 0$ for some $K(z)$ that does not depend on $M$.  From Proposition~\ref{prop:3.13}, we know that $\phi_M \in \mathcal{C}^2(0, M)$ satisfies 
\beq \label{eq:3.44}
-\lambda \phi_M(z)+ (\tilde{r}z-1)\phi_M'(z) +{1 \over 2}b^2 (1-\rho^2)z^2 \phi_M''(z)-m {(\phi_M'(z) )^2\over \phi_M''(z)}=0,
\endeq

\noi for $z \in (0,M)$, with the constant $m$ defined in \eqref{eq:2.14}.  After multiplying both sides of $\eqref{eq:3.44}$ by $\phi_M''(z) > 0$, we obtain

\beq \label{eq:3.45}
 P(\phi_M''(z)) := {1 \over 2}b^2 (1-\rho^2)z^2 (\phi_M''(z))^2+ (\tilde{r}z-1)\phi_M'(z) \phi_M''(z) -m (\phi_M'(z))^2>0, 
 \endeq

\noi for $z \in (0,M)$, since ${\phi}_M(z) > 0$ for $z \in (0,M)$. The polynomial $P$ has one positive and one negative root, $r_+(z)$ and $r_-(z)$, respectively. Since $\phi_M''(z) > 0$, it follows from \eqref{eq:3.45} that $\phi_M''(z)\geq r_+(z)= \gamma(z) \phi_M'(z) > 0$ for $z \in (0, M)$, where
\[
\gamma(z):=\frac{-(\tilde{r}z-1)-\sqrt{(\tilde{r}z-1)^2+2mb^2 (1-\rho^2)z^2}}{b^2 (1-\rho^2) z^2}<0.
\]
Since $\phi'_M(z) \leq C_2<0$ for $z < z_2$, we have that $\phi_M''(z)>K(z):=C_2 \gamma(z)$ for $z<z_2$.

 In the rest we will assume that $\mu > r + \sigma b \rho$. The case when $\mu \leq r + \sigma b \rho$ can be similarly handled. (Note that this condition merely changes the sign of the optimizer in the HJB equation for $\phi_M$.)

The function $\phi_M$ is a smooth solution of 

\[
\begin{split}
&  \la \, f = (\tilde r z - 1) \, f' + {1 \over 2} \, b^2 \, (1 - \rho^2) \, z^2 \, f'' + \min_{0 \leq \tilde {\pi} \leq L(z)} \left[ (\mu - r - \sigma b \rho) \, \tilde \pi \, f' + {1 \over 2} \, \sigma^2 \, \tilde \pi^2 \, f'' \right], \cr
&  f(z_1) = \phi_M(z_1) \hbox{ and } f(z_2)=\phi_M(z_2) ,
\end{split}
\]

\noindent in which $L(z) = -{\mu -r-b \rho \sigma \over \sigma^2}{C_1 \over C_2 \gamma(z)}$.

Next, by repeating the proof of Proposition~\ref{prop:3.14} after replacing $F$ in \eqref{eq:3.41} with

\[
\begin{split}
F(z, u(z), u'(z), u''(z)) &= \lambda u(z) - (\tilde{r}z-1) u'(z) - {1 \over 2}b^2 (1-\rho^2) z^2 u''(z) \cr
& \quad -\min_{0 \leq \tilde{\pi} \leq L(z)}\left[(\mu-r-\sigma\, b\, \rho)\tilde{\pi}u'(z) + {1 \over 2}\sigma^2 \tilde{\pi}^2 u''(z) \right],
\end{split}
\]

\noi we obtain that the function $\tilde{\phi}$ is a viscosity solution of
\beq \label{eq:3.46}
\begin{split}
&  \la \, f = (\tilde r z - 1) \, f' + {1 \over 2} \, b^2 \, (1 - \rho^2) \, z^2 \, f'' + \min_{0 \leq \tilde \pi \leq L(z)} \left[ (\mu - r - \sigma b \rho) \, \tilde \pi \, f' + {1 \over 2} \, \sigma^2 \, \tilde \pi^2 \, f'' \right], \cr
&  f(z_1)=\tilde{\phi}(z_1)  \hbox{ and }  f(z_2) = \tilde{\phi}(z_2).
\end{split}
\endeq

On the other hand, \eqref{eq:3.46} has a unique viscosity solution; see Ishii and Lions (1990).  In addition, \eqref{eq:3.46} has a unique {\it smooth} solution; see Duffie et al.\ (1997, page 767) or Krylov (1987).  Because the choices of $z_1$ and $z_2$ are arbitrary, we conclude that $\tilde{\phi} \in \mathcal{C}^2(\R_+)$.  Since we have already proved in Proposition~\ref{prop:3.14} that $\tilde{\phi}$ is a viscosity solution of \eqref{eq:2.5}, we can immediately conclude that $\tilde{\phi}$ is classical solution of \eqref{eq:2.5}. \end{proof}

{\prop \label{prop:3.16}Let $\phi$ be as in $\eqref{eq:2.4}$ and $\tilde{\phi}$ as in $\eqref{eq:3.40}$. Then, $\phi = \tilde{\phi}$ on $\rp$. Moreover, $\phi$ is the unique classical solution of $\eqref{eq:2.5}$ with optimal investment strategy given in $(2.6)$, and $\phi$ is strictly decreasing and strictly convex.}

\pf By using a verification lemma similar to Lemma~\ref{lem:3.12}, we can show that if there exists a smooth solution to the HJB equation in \eqref{eq:2.5}, then it equals $\phi$ with optimal investment strategy given in (2.6).  But, as we have shown in Proposition~\ref{prop:3.15}, $\tilde{\phi}$ is a classical solution of \eqref{eq:2.5}; therefore, the first claim follows. The convexity of $\phi$ follows since $\tilde{\phi}$, being the upper  envelope of convex functions, is convex. Since $\phi$ satisfies \eqref{eq:2.5} it is strictly convex.
That $\phi$ is strictly decreasing follows from the fact that each $\phi_M$ is decreasing on $[0, M]$ and that $\phi$ is strictly convex on $\rp$.   \end{proof}

\subsection{Proof of Theorem~\ref{thm:2.1}}\label{sec:3.4}

In this section, we complete our long series of propositions with a brief proof of Theorem~\ref{thm:2.1}.

{\prop \label{prop:3.17} Define $\tilde \psi$ on $\rp^2$ by $\tilde \psi(w, c) = \phi(w/c)$.  Then, $\psi = \tilde \psi$ on $\rp^2$.  Moreover, $\psi$ is the unique classical solution of $\eqref{eq:2.2}$ with optimal investment strategy given in \eqref{eq:moptinvst}.  Also, $\psi$ is strictly decreasing and strictly convex with respect to $w$ and strictly increasing with respect to $c$.}

\pf  By using a verification lemma similar to Lemma~\ref{lem:3.12}, we can show that if there exists a smooth solution to the HJB equation in \eqref{eq:2.2}, then it equals $\psi$ with optimal investment strategy given in \eqref{eq:moptinvst}. It is straightforward to show that $\tilde \psi$ solves \eqref{eq:2.2}; therefore, the claim follows.  Next, $\psi$ is strictly decreasing and strictly convex with respect to $w$ because $\phi$ is strictly decreasing and strictly convex on $\rp$.  Finally, $\psi$ is strictly increasing with respect to $c$ because $\phi$ is strictly decreasing on $\rp$.  \end{proof}

\section{Summary and Conclusions}\label{sec:4}

We studied three important problems of optimal control and showed how their value functions are related.  We first showed that our value functions are viscosity solutions of the corresponding HJB (in)equalities and later upgraded the regularity of the solutions by using the fact that the functions we analyzed are known to be value functions rather than merely solutions of HJB equations.  As a result, we used both probabilistic arguments (or arguments from control theory) and differential equations to show this further regularity.  We used a wide variety of techniques to prove these properties, including methods from viscosity solutions (Propositions \ref{prop:3.3}, \ref{prop:3.6}, \ref{prop:3.8}, \ref{prop:3.14}, and \ref{prop:3.15}), optimal stopping theory (Proposition~\ref{prop:3.9}), probabilistic arguments (Proposition~\ref{prop:3.10}), and verification lemmas (Lemma \ref{lem:3.12} and Propositions \ref{prop:3.13}, \ref{prop:3.16}, and \ref{prop:3.17}).

{\footnotesize
\section*{References}


\smallskip \noindent \hangindent 20 pt Bayraktar, E. and V. R. Young (2007a), Minimizing the probability of lifetime ruin under borrowing constraints, {\it Insurance: Mathematics and Economics}, 41: 196-221.

\smallskip \noindent \hangindent 20 pt Bayraktar, E. and V. R. Young (2007b), Correspondence between lifetime minimum wealth and utility of consumption, {\it Finance and Stochastics}, 11 (2): 213-236.

\smallskip \noindent \hangindent 20 pt Bayraktar, E. and V. R. Young (2008), Mutual fund theorems when minimizing the probability of lifetime ruin, {\it Finance Research Letters}, 5 (2): 69-78.

\smallskip \noindent \hangindent 20 pt Bayraktar, E., Q. Song, and J. Yang (2010), On the Continuity of Stochastic Exit Time Control Problems, to appear in {\it Stochastic Analysis and Applications}.

\smallskip \noindent \hangindent 20 pt Browne, S. (1995), Optimal investment policies for a firm with a random risk process: Exponential utility and minimizing the probability of ruin, {\it Mathematics of Operations Research}, 20 (4): 937-958.

\smallskip \noindent \hangindent 20 pt Buckdahn R. and J. Li (2009), Probabilistic interpretation for systems of Isaacs equations with two reflecting barriers, {\it Nonlinear Differential Equations and Applications}, 16: 381-420.

\smallskip \noindent \hangindent 20 pt Crandall, M., H. Ishii, and P.-L. Lions (1992), User's guide to viscosity solutions of second order partial differential equations, {\it Bulletin of American Mathematical Society}, 27 (1): 1-67.

\smallskip \noindent \hangindent 20 pt Duffie, D. and T. Zariphopoulou (1993), Optimal investment with undiversifiable income risk, {\it Mathematical Finance}, 3: 135-148.

\smallskip \noindent \hangindent 20 pt Duffie, D., W. Fleming, M. Soner, and T. Zariphopoulou (1997), Hedging in incomplete markets with HARA utility, {\it Journal of Economic Dynamics and Control}, 21: 753-782.


\smallskip \noindent \hangindent 20 pt Fleming, W. H. and P. E. Souganidis (1989), On the Existence of Value Functions of Two-Player, Zero-Sum Stochastic Differential Games, {\it Indiana University Mathematics Journal}, 38 (2): 293-314.


\smallskip \noindent \hangindent 20 pt Ishii, H. and P.-L. Lions (1990), Viscosity solutions of fully nonlinear second-order elliptic partial differential equations, {\it Journal of Differential Equations}, 83 (1): 26-78. 

\smallskip \noindent \hangindent 20 pt Karatzas, I. and S. Shreve (1998), {\it Methods of Mathematical Finance}, Springer-Verlag, New York.

\smallskip \noindent \hangindent 20 pt Karatzas, I. and W. D. Sudderth (2001), The controller-and-stopper game for a linear diffusion, {\it Annals of Probability}, 29: 1111-1127.

\smallskip \noindent \hangindent 20 pt Karatzas, I. and M. Zamfirescu (2006a), Martingale approach to stochastic control with discretionary stopping, {\it Applied Mathematics and Optimization}, 53: 163-184.

\smallskip \noindent \hangindent 20 pt Karatzas, I. and M. Zamfirescu (2008), Martingale approach to stochastic differential games of control and stopping, {\it Annals of Probability}, 36 (4), 1495-1527.



Karlin, S. and Taylor, H. M. (1981). .
Academic Press Inc. [

\smallskip \noindent \hangindent 20 pt Karlin, S., and H. M. Taylor  (1981), {\it A Second Course in Stochastic Processes}, Academic Press, New York.

\smallskip \noindent \hangindent 20 pt Lions, P.-L. (1983), Optimal control of diffusion processes and Hamilton-Jacobi-Bellman equations, Part 2: viscosity solutions and uniqueness, {\it Communications in Partial Differential Equations}, 8 (11): 1229-1276.

\smallskip \noindent \hangindent 20 pt Milevsky, M. A., K. Ho, and C. Robinson (1997), Asset allocation via the conditional first exit time or how to avoid outliving your money, {\it Review of Quantitative Finance and Accounting}, 9 (1): 53-70.

\smallskip \noindent \hangindent 20 pt Milevsky, M. A., K. S. Moore, and V. R. Young (2006), Asset allocation and annuity-purchase strategies to minimize the probability of financial ruin, {\it Mathematical Finance}, 16 (4): 647-671.

\smallskip \noindent \hangindent 20 pt Milevsky, M. A. and C. Robinson (2000), Self-annuitization and ruin in retirement, with discussion, {\it North American Actuarial Journal}, 4 (4): 112-129.

\smallskip \noindent \hangindent 20 pt Pham, H. (1998), Optimal stopping of controlled jump diffusion process: a viscosity solution approach, {\it Journal of Mathematical Systems, Estimation, and Control}, 8 (1): 1-27.

\smallskip \noindent \hangindent 20 pt Pham, H.  (2009), {\it Continuous-time stochastic control and optimization with financial applications}, Stochastic Modelling and Applied Probability, Vol. 61, Springer, Berlin.

\smallskip \noindent \hangindent 20 pt Rogers, L. C. G. and Williams, David (2000), {\it Diffusions, {M}arkov processes, and martingales. {V}ol. 2}, Cambridge Mathematical Library, Cambridge University Press, Cambridge. Reprint of the second (1994) edition

\smallskip \noindent \hangindent 20 pt  Shreve, S.  and H. M. Soner  (1994), Optimal investment and consumption with transaction costs, {\it Annals of Applied Probability}, 4 (3), 609-692.

\smallskip \noindent \hangindent 20 pt Touzi, N. (2002), Stochastic control and application to finance, {\it Scuola Normale Superiore, Pisa. Special Research Semester on Financial Mathematics, Lecture Notes}, available at http://www.cmap.polytechnique.fr/$\sim$touzi/pise02.pdf


\smallskip \noindent \hangindent 20 pt Young, V. R. (2004), Optimal investment strategy to minimize the probability of lifetime ruin, {\it North American Actuarial Journal}, 8 (4): 105-126.

\smallskip \noindent \hangindent 20 pt Zariphopoulou, T. (1994), Consumption-investment models with constraints, {\it SIAM Journal on Control and Optimization}, 32 (1): 59-85.
}

\end{document}